\documentclass[aps,prl,twocolumn,groupedaddress,nofootinbib]{revtex4-1}
\usepackage{amsmath}
\usepackage{amssymb}
\usepackage{graphicx,subfigure}
\usepackage{mathrsfs}
\usepackage{ntheorem}
\usepackage[T1]{fontenc}
\usepackage{bm}
\usepackage{subfigure}
\usepackage[bookmarks=false]{hyperref}
\usepackage{color}
\hypersetup{colorlinks=true,citecolor=blue,
linkcolor=red,urlcolor=blue,pdfstartview=FitH,
bookmarksopen=true}

\newcommand{\ket}[1]{|#1\rangle}
\newcommand{\bra}[1]{\langle #1|}
\newcommand{\inp}[2]{\langle #1|#2\rangle}
\newcommand{\inpp}[2]{\prec #1,#2 \succ}
\newcommand{\exppt}[1]{\prec #1 \succ}

\newcommand{\tr}{\mathrm{tr}}
\newcommand{\ud}{\mathrm{d}}

\newcommand{\abs}[1]{\lvert #1\rvert}

\def\CC{{\rm\kern.24em \vrule width.04em height1.46ex depth-.07ex \kern-.30em C}}
\def\RR{{\rm\kern.24em \vrule width.04em height1.46ex depth-.07ex
\kern-.30em R}}
\def\P{{\rm I\kern-.25em P}}

\begin{document}
\title{Quantum Geometric Tensor in $\mathcal{PT}$-Symmetric Quantum Mechanics}

\author{Da-Jian Zhang}
\affiliation{Department of Physics, National University of Singapore, Singapore 117542}
\author{Qing-hai Wang}
\affiliation{Department of Physics, National University of Singapore, Singapore 117542}
\author{Jiangbin Gong}
\email{phygj@nus.edu.sg}
\affiliation{Department of Physics, National University of Singapore, Singapore 117542}
\date{\today}

\begin{abstract}
A series of geometric concepts are formulated for $\mathcal{PT}$-symmetric quantum mechanics and they are further unified into one entity, i.e., an extended quantum geometric tensor (QGT). The imaginary part of the extended QGT gives a Berry curvature whereas the real part induces a metric tensor on system's parameter manifold. This results in a unified conceptual framework to understand and explore physical properties of $\mathcal{PT}$-symmetric systems from a geometric perspective. To illustrate the usefulness of the extended QGT, we show how its real part, i.e., the metric tensor, can be exploited as a tool to detect quantum phase transitions as well as spontaneous $\mathcal{PT}$-symmetry breaking in $\mathcal{PT}$-symmetric systems.
\end{abstract}

\maketitle

Given a family of Hamiltonians depending smoothly on a manifold of parameters, e.g., external field  strengths, a problem of great importance is how to characterize geometric aspects of the eigenstates of the Hamiltonians. In standard quantum mechanics (QM) where Hamiltonians are Hermitian operators, the solution to the problem is to use the quantum geometric tensor (QGT), of which the imaginary part determines the Berry curvature \cite{1984Berry45} and the real part induces a Riemannian metric tensor \cite{1980Provost289} on the manifold. The QGT has played an indispensable role in various frontier topics of quantum computation, quantum information, and condensed-matter physics \cite{1989Shapere, 2003Bohm}, where both its imaginary and real parts serve as versatile tools.

Since the pioneering work of Bender and Boettcher \cite{1998Bender5243}, however, it has been realized that
Hamiltonians can be non-Hermitian but still possess real spectra due to parity-time reversal ($\mathcal{PT}$) symmetry. This has led to a complex extension of standard QM called $\mathcal{PT}$-symmetric QM ($\mathcal{PT}$QM) \cite{2002Bender270401,2002Mostafazadeh205}, with one main conceptual advance being the introduction of a nontrivial inner-product metric to define its Hilbert space.
 Over the past decade, $\mathcal{PT}$-symmetric systems have been experimentally realized by spatially engineering gain-loss structures \cite{2018El-Ganainy11}, thus further boosting
$\mathcal{PT}$QM as an important research area.
A current stream of development is towards extensive studies of physical properties, especially topological properties, of $\mathcal{PT}$-symmetric systems \cite{
2017Ashida15791,2017Kawabata190401,
2017Weimann433,2017Menke174506a,
2018Kawabata85116,2018Lourenco85126,2018Shen146402,
2018Yao136802,2018Gong31079}.

Unfortunately, a systematic geometric concept like the QGT is still elusive in $\mathcal{PT}$QM. Without such a concept, it is difficult to extend geometric understandings, such as those of quantum phase transitions (QPTs) \cite{2005Carollo157203,2006Zhu77206,
2006Zanardi31123,2007Zanardi100603,2007Venuti95701,2007Cozzini14439,
2007Cozzini104420,2007Zanardi32109}, to $\mathcal{PT}$-symmetric systems. On the other hand,
the interplay between the nontrivial inner-product metric and geometric aspects of $\mathcal{PT}$QM is rarely  understood to date.
In particular, in the course of varying parameters of a $\mathcal{PT}$-symmetric system, the inner-product metric varies as well \cite{2013Gong485302}. How this feature impacts on previous geometric perspectives in standard QM (such as curvature and metric tensor) remains unknown.

In this Letter, we report the finding of an extended QGT in $\mathcal{PT}$QM, of which the imaginary part gives a Berry curvature whereas the real part induces a metric tensor on system's parameter manifold. This work thus gives a unified conceptual framework to understand and explore physical properties of $\mathcal{PT}$-symmetric systems from a geometric perspective, enabling one to readily extend known tools and methods in standard QM to $\mathcal{PT}$-symmetric systems.

To present our finding clearly, we recapitulate some fundamentals of $\mathcal{PT}$QM. Consider a system with $\mathcal{PT}$-symmetric Hamiltonian $H(\lambda)$, depending on some parameters denoted collectively by $\lambda$. Generally speaking, $\lambda$'s can be grouped into two regimes: a regime of unbroken $\mathcal{PT}$ symmetry where $H(\lambda)$ has a real spectrum and a complete set of eigenstates, and a regime of broken $\mathcal{PT}$ symmetry where at least part of the eigenvalues are complex.
In the unbroken regime, denoted as $M$, a consistent quantum theory can be built.
{Indeed, for any given $\lambda\in M$, there exists a positive definite operator $W(\lambda)$ such that $W(\lambda)H(\lambda)=H^\dag(\lambda)W(\lambda)$. This enables one to define a new inner product, $\inpp{\cdot}{\cdot}_\lambda := \bra{\cdot}W(\lambda)\ket{\cdot}$, referred to as the $\lambda$-dependent inner product. The physical Hilbert space, denoted as $\mathcal{H}(\lambda)$, is endowed with this new inner product. Accordingly, a Hermitian operator $X$ over $\mathcal{H}(\lambda)$, referred to as physical Hermitian operator, satisfies $\inpp{\cdot}{X\cdot}_\lambda=\inpp{X\cdot}{\cdot}_\lambda$. Evidently $H(\lambda)$ is a physical Hermitian operator. The theory is built consistently by identifying any observable with a physical Hermitian operator. It should be emphasized that
the choices of such $W(\lambda)$ are not unique, but the results of this Letter are irrespective of any specific choices of $W(\lambda)$.

With the above knowledge, we may now be able to establish the extended QGT. To do this, we first formulate a series of geometric concepts, such as Berry phase, Berry curvature, and metric tensor, and then propose a QGT to unify all of them.

\textit{First, we find a Berry phase}.
Consider an evolution where $\lambda_t\in M$ varies over a time interval $[0,\tau]$, i.e., $H(t)=H(\lambda_t)$. For this, $\mathcal{H}(\lambda_t)$ moves with time $t$, and the evolving state $\ket{\psi(t)}$ at time $t$
belongs to $\mathcal{H}(\lambda_t)$. The Schr\"{o}dinger-like equation for such an evolution is found to be \cite{2013Gong485302} ($\hbar=1$)
\begin{eqnarray}\label{eq:Sch}
i\partial_t\ket{\psi(t)}=\left[H(t)+iK(t)\right]\ket{\psi(t)},
\end{eqnarray}
where
$K(t):=-\textstyle{\frac{1}{2}}W^{-1}(\lambda_t)\partial_tW(\lambda_t)$
is a physical Hermitian operator, representing a gauge field necessary for unitarity. That is, the $\lambda$-dependent inner product of two arbitrary initial states is preserved during the evolution.
The form
of $K(t)$ has also been justified by others \cite{2018Mostafazadeh46022} and this Schr\"{o}dinger-like equation has already found a number of applications \cite{2015Deffner150601,2016Fring42128,2016Fring42114,2017Mead85001,
 2017Zeng31001,2018Wei12105,2018Wei12114}.

Suppose that $\lambda_t$ forms a closed curve $C$ in $M$, i.e., $\lambda_0=\lambda_\tau$, and moreover, it changes sufficiently slowly so that the adiabatic theorem applies \cite{1note-GQT}. Then, starting at the $n$-th eigenstate of $H(\lambda_0)$, the evolving state $\ket{\psi(t)}$ remains in the $n$-th instantaneous eigenstate of $H(\lambda_t)$:
\begin{eqnarray}\label{eq:evolving-state}
\ket{\psi(t)}=e^{i\alpha(t)}\ket{\Psi_n(\lambda_t)}.
\end{eqnarray}
Here, $\ket{\Psi_n(\lambda)}$ is the $n$-th eigenstate with the normalization condition $\inpp{\Psi_n(\lambda)}{\Psi_n(\lambda)}_\lambda=1$, i.e., $H(\lambda)\ket{\Psi_n(\lambda)}=E_n(\lambda)\ket{\Psi_n(\lambda)}$,
with $E_n(\lambda)$ denoting the associated eigenenergy,
and $E_m(\lambda)\neq E_n(\lambda)$ for all $m\neq n$ have been assumed. For later convenience, we let $\ket{\Phi_n(\lambda)}:=W(\lambda)\ket{\Psi_n(\lambda)}$.
Substituting Eq.~(\ref{eq:evolving-state}) into Eq.~(\ref{eq:Sch}) and contracting both sides of Eq.~(\ref{eq:Sch}) with $\bra{\Phi_n(\lambda_t)}$, we have
$\dot{\alpha}(t)=-E_n(\lambda_t)-
i\inpp{\Psi_n(\lambda_t)}{K(t)
\Psi_n(\lambda_t)}_{\lambda_t}
+\ i\inpp{\Psi_n(\lambda_t)}
{\dot{\Psi}_n(\lambda_t)}_{\lambda_t}$,
where
the dot denotes the time derivative.
Integrating this equation and noting that $\inpp{\Psi_n(\lambda_t)}{K(t)
\Psi_n(\lambda_t)}_{\lambda_t}=[\inpp{\dot{\Psi}_n(\lambda_t)}
{\Psi_n(\lambda_t)}_{\lambda_t}+\inpp{\Psi_n(\lambda_t)}
{\dot{\Psi}_n(\lambda_t)}_{\lambda_t}]/2$, we obtain
$\alpha(\tau)-\alpha(0)=\beta_n+\gamma_n$, with
$\beta_n:=-\int_0^\tau E_n(\lambda_t)\ud t$ and
$\gamma_n:=-\Im\int_0^t\inpp{\Psi_n(\lambda_t)}
{\dot{\Psi}_n(\lambda_t)}_{\lambda_t}\ud t$. Evidently $\beta_n$ is a dynamical phase. On the contrary, $\gamma_n$, as a phase obtained by removing the dynamical phase from the total phase change, is of geometric nature. To see this, we cast $\gamma_n$ in the form
\begin{eqnarray}\label{cp:BP}
\gamma_n=-\oint_C A_n,
\end{eqnarray}
with
$A_n:=\Im\inpp{\Psi_n(\lambda)}
{\partial_{\mu}\Psi_n(\lambda)}_{\lambda}\ud\lambda^\mu$
being a connection one-form. Here and henceforth, $\mu$'s label the components of $\lambda$, and the Einstein summation convention is assumed. Now, it is clear that $\gamma_n$ depends solely upon the closed curve $C$, thus representing a Berry phase.
It is interesting to note that $\gamma_n$ in Eq.~(\ref{cp:BP})
shares the same form with the seminal Berry phase \cite{1984Berry45}.
In passing, two previous studies on this subject \cite{2010Gong12103,2013Gong485302} did not observe that $\gamma_n$ can be expressed as one single line integral as in Eq.~(\ref{cp:BP}).

\textit{Second, we specify the associated Berry curvature.}  Using Eq.~(\ref{cp:BP}) and by Stokes' theorem, we have
\begin{eqnarray}\label{cp:BP-BC}
\gamma_n=-\int_S\Omega_n.
\end{eqnarray}
Here, $S$ is any surface enclosed by the curve $C$, and $\Omega_n:=\ud A_n$ is a two-form on the manifold $M$, representing a Berry curvature responsible for the appearance of $\gamma_n$, with $\ud$ denoting the exterior derivative on $M$.
To obtain
an explicit form of $\Omega_n$, let $A_{n,\mu}:=\Im\inpp{\Psi_n(\lambda)}
{\partial_{\mu}\Psi_n(\lambda)}_{\lambda}$. So, $A_n=A_{n,\mu}\ud\lambda^\mu$. Substituting it into $\Omega_n=\ud A_n$ yields
$\Omega_n=\partial_\mu A_{n,\nu}\ud\lambda^\mu\wedge\ud\lambda^\nu$, where
$\wedge$ denotes the wedge product. Using
the anti-commutativity of the wedge product, i.e., $\ud \lambda^\mu\wedge\ud \lambda^\nu=-\ud \lambda^\nu\wedge\ud \lambda^\mu$, we can rewrite this expression as
\begin{eqnarray}\label{eq:F}
\Omega_n=\textstyle{\frac{1}{2}}\left(\partial_\mu A_{n,\nu}-\partial_\nu A_{n,\mu}\right)\ud\lambda^\mu\wedge\ud\lambda^\nu.
\end{eqnarray}
Hence, the components of $\Omega_n$ read
\begin{eqnarray}\label{eq:F-com}
\Omega_{n,\mu\nu}=\textstyle{\frac{1}{2}}\left(\partial_\mu A_{n,\nu}-\partial_\nu A_{n,\mu}\right),
\end{eqnarray}
i.e., $\Omega_n=\Omega_{n,\mu\nu}\ud\lambda^\mu\wedge\ud\lambda^\nu$.
Inserting $A_{n,\mu}=\Im\inpp{\Psi_n(\lambda)}
{\partial_{\mu}\Psi_n(\lambda)}_{\lambda}$ into Eq.~(\ref{eq:F-com}) and noting that $\inpp{\Psi_n(\lambda)}
{\partial_{\mu}\Psi_n(\lambda)}_{\lambda}=
\inp{\Phi_n(\lambda)}{\partial_{\mu}\Psi_n(\lambda)}$,
we arrive at the explicit form:
\begin{eqnarray}\label{cp:BCU-component}
\Omega_{n,\mu\nu}=\textstyle{\frac{1}{2}}\Im\left[\inp{\partial_{\mu}
\Phi_n(\lambda)}
{\partial_{\nu}\Psi_n(\lambda)}
+\inp{\partial_{\mu}\Psi_n(\lambda)}
{\partial_{\nu}\Phi_n(\lambda)}\right].\nonumber\\
\end{eqnarray}
Here, $\Im\inp{\partial_{\nu}\Phi_n(\lambda)}
{\partial_{\mu}\Psi_n(\lambda)}=-
\Im\inp{\partial_{\mu}\Psi_n(\lambda)}
{\partial_{\nu}\Phi_n(\lambda)}$ has been used. Similar to the seminal one \cite{1983Simon2167}, the Berry curvature (\ref{cp:BCU-component}) is a real anti-symmetric tensor, i.e., $\Omega_{n,\mu\nu}=\Omega_{n,\mu\nu}^*$ and $\Omega_{n,\mu\nu}=-\Omega_{n,\nu\mu}$.

\textit{Third, we formulate the concept of metric tensor}. We start with the introduction of a density operator. It is easy to see that $\rho_n(\lambda):=
\ket{\Psi_n(\lambda)}\bra{\Phi_n(\lambda)}$
is a positive operator over $\mathcal{H}(\lambda)$ satisfying $\tr[\rho_n(\lambda)]=1$ and $\rho_n^2(\lambda)=\rho_n(\lambda)$; it fulfills the conditions of being a density operator for a pure state. So, $\rho_n(\lambda)$ can be seen as the density operator associated to $\ket{\Psi_n(\lambda)}$.
We then propose a formula for the fidelity between  $\rho_n(\lambda)$ and $\rho_n(\lambda+\delta\lambda)$.
It reads
$F(\rho_n(\lambda),\rho_n(\lambda+\delta\lambda)):=\tr\sqrt{
\abs{\rho_n^{1/2}(\lambda)\rho_n(\lambda+\delta\lambda)
\rho_n^{1/2}(\lambda)}}$. Here, for an operator $X$, $\abs{X}:=\sqrt{X^\ddagger X}$, with $X^\ddagger$ being the Hermitian conjugate of $X$ w.r.t. the $\lambda$-dependent inner product. This formula is almost of the same form as that in standard QM \cite{2010Nielsen}. Inserting $\rho_n(\lambda)=
\ket{\Psi_n(\lambda)}\bra{\Phi_n(\lambda)}$ and $\rho_n(\lambda+
\delta\lambda)=
\ket{\Psi_n(\lambda+\delta\lambda)}\bra{\Phi_n(\lambda+\delta
\lambda)}$
into this formula yields
\begin{eqnarray}\label{df:fidelity}
&&F(\rho_n(\lambda),\rho_n(\lambda+\delta\lambda))\nonumber\\
&=&\sqrt{\abs{\inp{\Phi_n(\lambda+\delta\lambda)}{\Psi_n(\lambda)}
\inp{\Phi_n(\lambda)}{\Psi_n(\lambda+\delta\lambda)}
}}.
\end{eqnarray}
We now specify the metric tensor. In the spirit of Bures distance \cite{1980Provost289}, the distance element between $\rho_n(\lambda)$ and $\rho_n(\lambda+\delta\lambda)$ can be defined as
$\ud s^2:=2[1-F(\rho_n(\lambda),\rho_n(\lambda+\delta\lambda))]$.
Substituting Eq.~(\ref{df:fidelity}) into this expression and using Taylor-series expansions of $\ket{\Psi_n(\lambda+\delta\lambda)}$
and $\ket{\Phi_n(\lambda+\delta\lambda)}$, we obtain, up to second order, $\ud s^2=g_{n,\mu\nu}\ud\lambda^\mu\ud\lambda^\nu$, with \cite{1QGT_SM}
\begin{eqnarray}\label{cp:metric-tensor}
g_{n,\mu\nu}=&&\textstyle{\frac{1}{2}}\Re[\inp{\partial_{\mu}
\Phi_n(\lambda)}{\partial_{\nu}\Psi_n(\lambda)}
-\inp{\partial_{\mu}\Phi_n(\lambda)}{\Psi_n(\lambda)}
\times\nonumber\\
&&\inp{\Phi_n(\lambda)}{\partial_{\nu}\Psi_n(\lambda)}
+\textrm{``terms $\Phi\leftrightarrow\Psi$''}].
\end{eqnarray}
Here, $\textrm{``terms $\Phi\leftrightarrow\Psi$''}$ stands for $\inp{\partial_{\mu}
\Psi_n(\lambda)}{\partial_{\nu}\Phi_n(\lambda)}
-\inp{\partial_{\mu}\Psi_n(\lambda)}{\Phi_n(\lambda)}
\inp{\Psi_n(\lambda)}{\partial_{\nu}\Phi_n(\lambda)}$.
Equation (\ref{cp:metric-tensor}) gives the desired metric tensor. Like the seminal metric tensor \cite{1980Provost289}, it is a real symmetric tensor, i.e., $g_{n,\mu\nu}=g_{n,\mu\nu}^*$ and  $g_{n,\mu\nu}=g_{n,\nu\mu}$.

\textit{Finally, we are ready to present the extended QGT.} It reads
\begin{eqnarray}\label{cp:QGT}
Q_{n,\mu\nu}:=&&\textstyle{\frac{1}{2}}[\inp{\partial_{\mu}
\Phi_n(\lambda)}{\partial_{\nu}\Psi_n(\lambda)}
-\inp{\partial_{\mu}\Phi_n(\lambda)}{\Psi_n(\lambda)}
\times\nonumber\\
&&\inp{\Phi_n(\lambda)}{\partial_{\nu}\Psi_n(\lambda)}
+\textrm{``terms $\Phi\leftrightarrow\Psi$''}].
\end{eqnarray}
Analogous to the seminal one \cite{1980Provost289}, the extended QGT is a complex Hermitian tensor, i.e., $Q_{n,\mu\nu}=Q_{n,\nu\mu}^*$.
Moreover, it is independent of any specific choices of $W(\lambda)$ as long as $W(\lambda)$ satisfies $W(\lambda)H(\lambda)=H^\dag(\lambda)W(\lambda)$ \cite{1QGT_SM}.
On the other hand, the extended QGT
unifies all the geometric concepts formulated in the previous paragraphs. To see this, we examine its imaginary and real parts, respectively.
From the equalities $\inp{\partial_\mu\Phi_n(\lambda)}{\Psi_n(\lambda)}=
-\inp{\partial_\mu\Psi_n(\lambda)}{\Phi_n(\lambda)}^*$
and $\inp{\Phi_n(\lambda)}{\partial_\nu\Psi_n(\lambda)}=
-\inp{\Psi_n(\lambda)}{\partial_\nu\Phi_n(\lambda)}^*$,
we deduce that $-\inp{\partial_\mu\Phi_n(\lambda)}{\Psi_n(\lambda)}
\inp{\Phi_n(\lambda)}{\partial_\nu\Psi_n(\lambda)}-
\inp{\partial_\mu\Psi_n(\lambda)}{\Phi_n(\lambda)}
\inp{\Psi_n(\lambda)}{\partial_\nu\Phi_n(\lambda)}$, i.e., a term appearing in Eq.~(\ref{cp:QGT}),
is real, which leads to
$\Im \left[Q_{n,\mu\nu}\right]=\Im(\inp{\partial_{\mu}
\Phi_n(\lambda)}{\partial_{\nu}\Psi_n(\lambda)}+
\inp{\partial_{\mu}
\Psi_n(\lambda)}{\partial_{\nu}\Phi_n(\lambda)})/2$. That is,
\begin{eqnarray}\label{re:Im}
\Im \left[Q_{n,\mu\nu}\right]=\Omega_{n,\mu\nu}.
\end{eqnarray}
Further, comparing Eq.~(\ref{cp:metric-tensor}) with Eq.~(\ref{cp:QGT}) gives
\begin{eqnarray}\label{re:Re}
\Re \left[Q_{n,\mu\nu}\right]=g_{n,\mu\nu}.
\end{eqnarray}
Hence, we arrive at the claimed unification:
The imaginary part of the QGT gives the Berry curvature (\ref{cp:BCU-component}) and thus further determines the Berry phase (\ref{cp:BP-BC}), whereas the real part induces the metric tensor (\ref{cp:metric-tensor}) and thereby further determines the fidelity (\ref{df:fidelity}).

So far, we have established the extended QGT. To illustrate its usefulness, we show how its real part, i.e., the metric tensor, can be used to detect quantum criticality of $\mathcal{PT}$-symmetric systems.

As we know, a system remains in its ground state (GS) at zero temperature, irrespective of system's parameters. Accordingly, the manifold of parameters can be partitioned into regions characterized by the fact that inside them the GS can move ``adiabatically'' from one point to the other and no singularities in expectation values of any observables are encountered. The boundaries between these ``regular'' regions, referred to as critical points, are in turn with abrupt changes in the GS, resulting in singular behaviors of some observables. Such abrupt changes are due to the presence of points of degeneracy of the GS. Here, one should distinguish two types of degeneracy. One is the level crossing or avoided crossing between the GS and excited states, as in QPTs \cite{1999Sachdev}.
The other is the spectral coalescence because of spontaneous $\mathcal{PT}$ symmetry breaking \cite{2013Bender173}.
These two types of degeneracy lead to two kinds of critical points, referred to as the QPT point and the $\mathcal{PT}$ phase transition ($\mathcal{PT}$PT) point, respectively.

To reveal critical points, we resort to the metric tensor (\ref{cp:metric-tensor}). The metric tensor associated to the GS can be expressed as \cite{1QGT_SM}
\begin{widetext}
\begin{eqnarray}\label{eq:metric-tensor-GS}
g_{0,\mu\nu}=\Re\sum_{n\neq 0}
\frac{\bra{\Phi_0(\lambda)}\partial_\mu H(\lambda)\ket{\Psi_n(\lambda)}\bra{\Phi_n(\lambda)}
\partial_\nu H(\lambda)\ket{\Psi_0(\lambda)}+\bra{\Phi_n(\lambda)}
\partial_\mu
H(\lambda)\ket{\Psi_0(\lambda)}\bra{\Phi_0(\lambda)}
\partial_\nu
H(\lambda)\ket{\Psi_n(\lambda)}}
{2\left[E_0(\lambda)-E_n(\lambda)\right]^2}.
\end{eqnarray}
\end{widetext}
Here, the subscript $0$ labels the GS. Clearly, the degeneracy of the GS at critical points amounts to a vanishing denominator in Eq.~(\ref{eq:metric-tensor-GS}). This may
break down the analyticity of the metric tensor. For this reason, the singularities of the metric tensor can serve as  signatures of presence of critical points, which makes the metric tensor useful for detecting regions of criticality.

To gain more physical insight, we link the metric tensor to fluctuations of expectation values of some observables. To proceed, we introduce the operator $O_\mu:=i\sum_n\ket{\partial_\mu\Psi_n(\lambda)}
\bra{\Phi_n(\lambda)}$. Since $\inp{\Phi_m(\lambda)}{\Psi_n(\lambda)}=\delta_{mn}$ and $\sum_{n}\ket{\Psi_n(\lambda)}\bra{\Phi_n(\lambda)}=I$
\cite{1QGT_SM}, $O_\mu$ satisfies $i\ket{\partial_\mu\Psi_n(\lambda)}=O_\mu\ket{\Psi_n(\lambda)}$
and
$i\ket{\partial_\mu\Phi_n(\lambda)}=O_\mu^\dagger
\ket{\Phi_n(\lambda)}$. Here, $\delta_{mn}$ denotes the Kronecker delta symbol and $I$ is the identity operator. Moreover, $O_\mu$ can be decomposed as
$O_\mu=O_{A,\mu}+iO_{B,\mu}$, with $O_{A,\mu}$ and $O_{B,\mu}$ being two physical Hermitian operators. It can be shown \cite{1QGT_SM} that $O_{B,\mu}=-W^{-1}(\lambda)\partial_\mu W(\lambda)/2$, which is essentially the gauge field $K(t)$.
 Using these facts, we can cast the metric tensor in the form \cite{1QGT_SM}
\begin{eqnarray}\label{eq:semi-covariance-matrix-form}
g_{0,\mu\nu}=\textstyle{\frac{1}{2}}
\left(\exppt{\{\bar{O}_{A,\mu},\bar{O}_{A,\nu}\}}_\lambda
-\exppt{\{\bar{O}_{B,\mu},
\bar{O}_{B,\nu}\}}_\lambda\right).\nonumber\\
\end{eqnarray}
Here, for an operator $X$, $\bar{X}:=X-\exppt{X}_\lambda$ and $\exppt{X}_\lambda:=\inpp{\Psi_0(\lambda)}{X
\Psi_0(\lambda)}_\lambda$, i.e., the expectation value of $X$ in the GS. Letting $O_A:=O_{A,\mu}\ud\lambda^\mu$ and
$O_B:=O_{B,\mu}\ud\lambda^\mu$ and using Eq.~(\ref{eq:semi-covariance-matrix-form}), we cast the distance element in the form
\begin{eqnarray}\label{eq:metric-tensor-variance}
\ud s^2
=\exppt{\bar{O}_A^{2}}_\lambda-
\exppt{\bar{O}_B^{2}}_\lambda,
\end{eqnarray}
representing the difference between the variance of $O_A$ and that of $O_B$. Hence, the singular behavior of the metric tensor at critical points amounts to the fact that either the variance of $O_A$ or that of $O_B$ gets very large, possibly divergent, there. If we interpret $O_A$ and $O_B$ as two ``order parameters,'' the singular behavior of the metric tensor can be seen as that of susceptibilities of these two ``order parameters.''

\begin{figure}
  \centering
  \subfigure{
  \includegraphics[width=.22\textwidth]{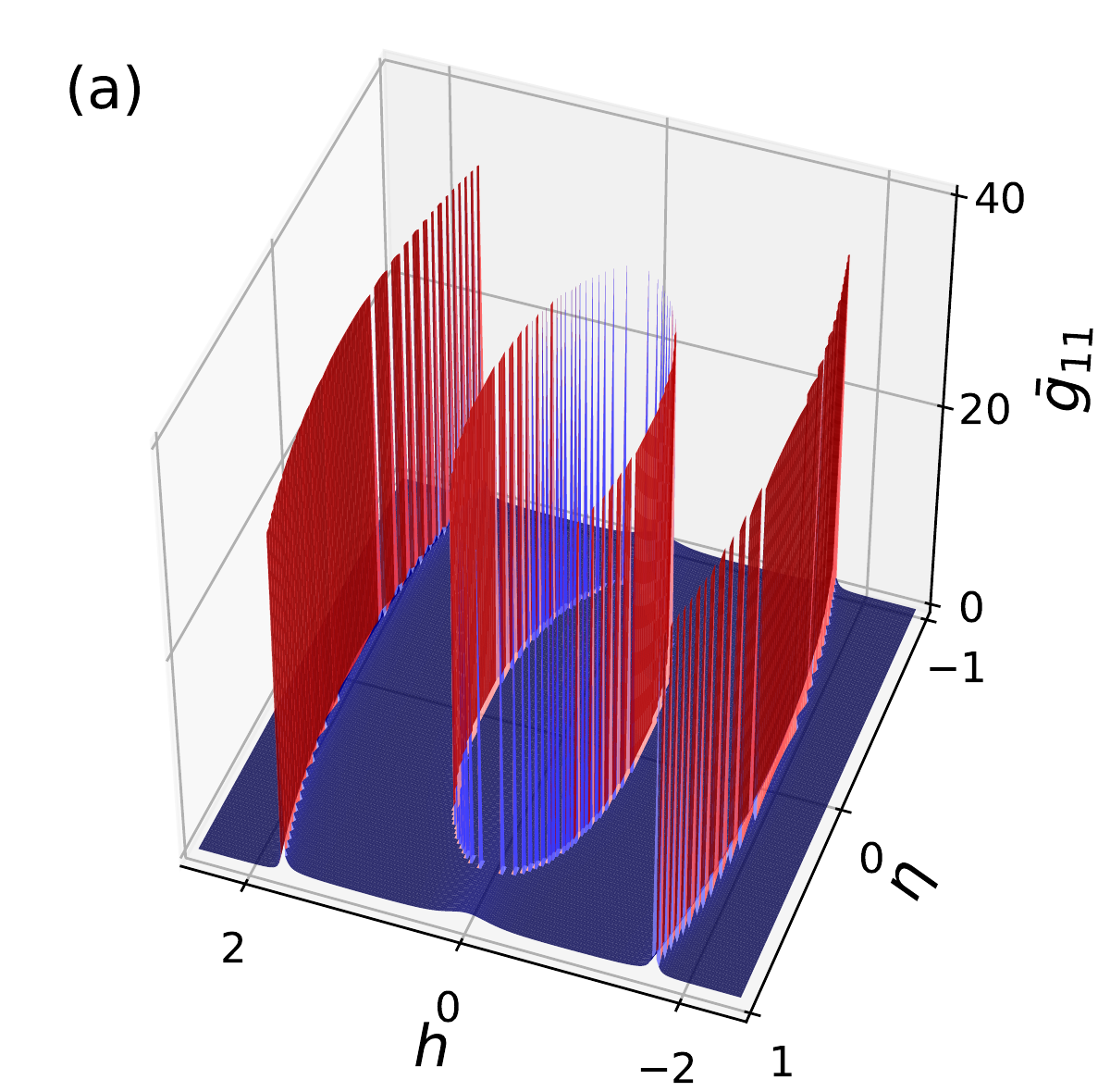}\label{fig1}}
  \subfigure{
  \includegraphics[width=.22\textwidth]{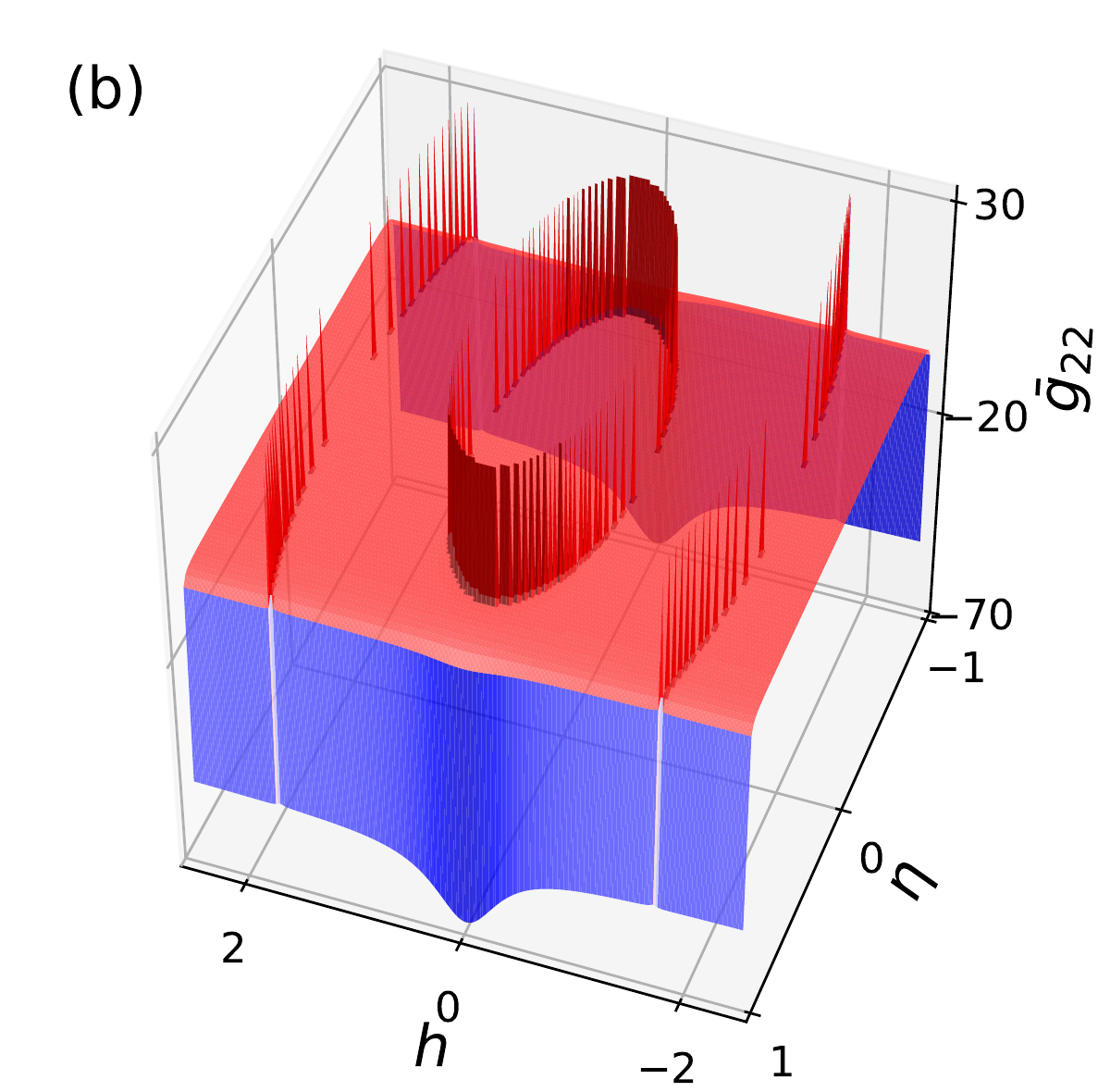}\label{fig2}}\\
  \subfigure{
  \includegraphics[width=.22\textwidth]{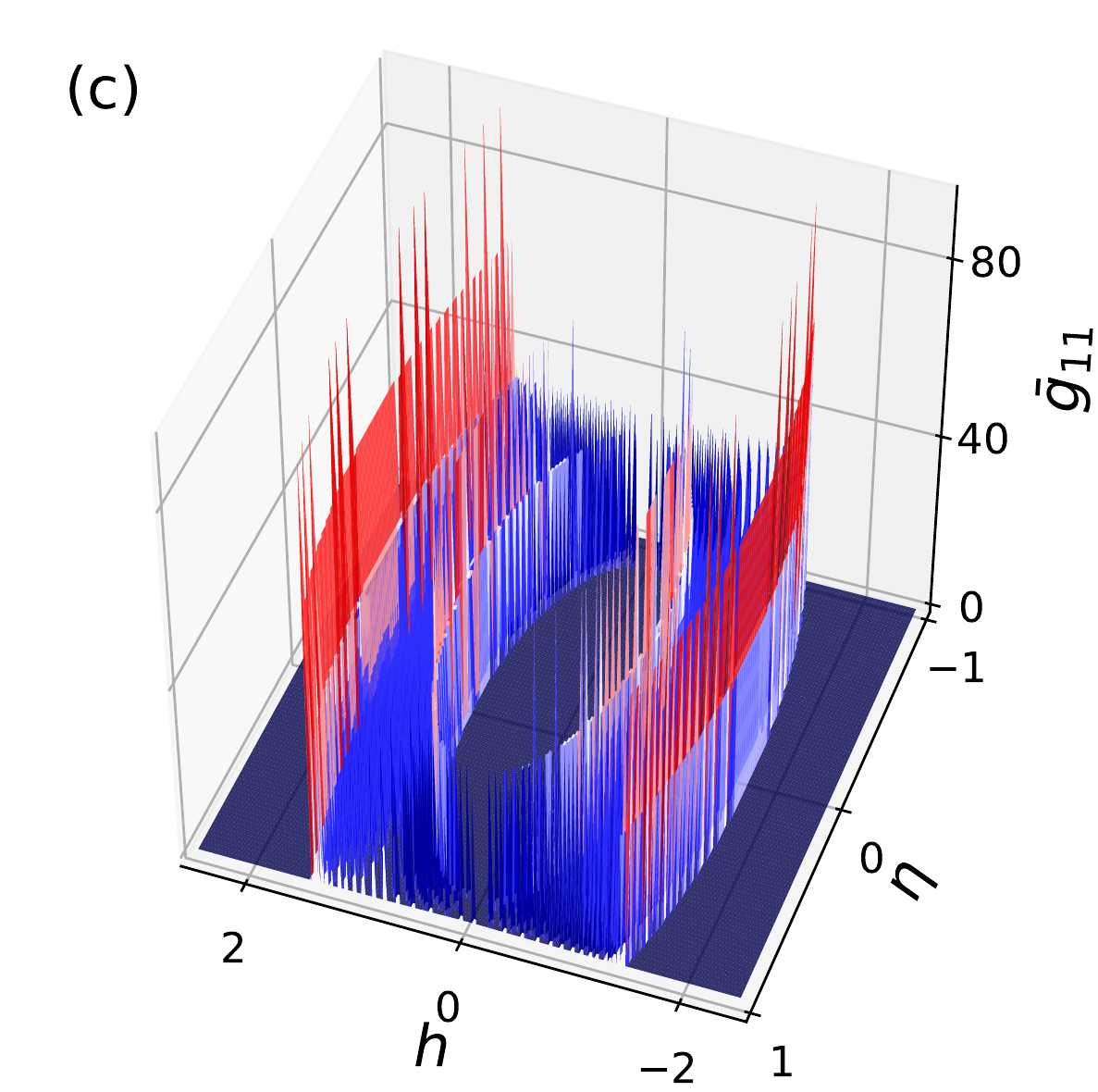}\label{fig3}}
  \subfigure{
  \includegraphics[width=.22\textwidth]{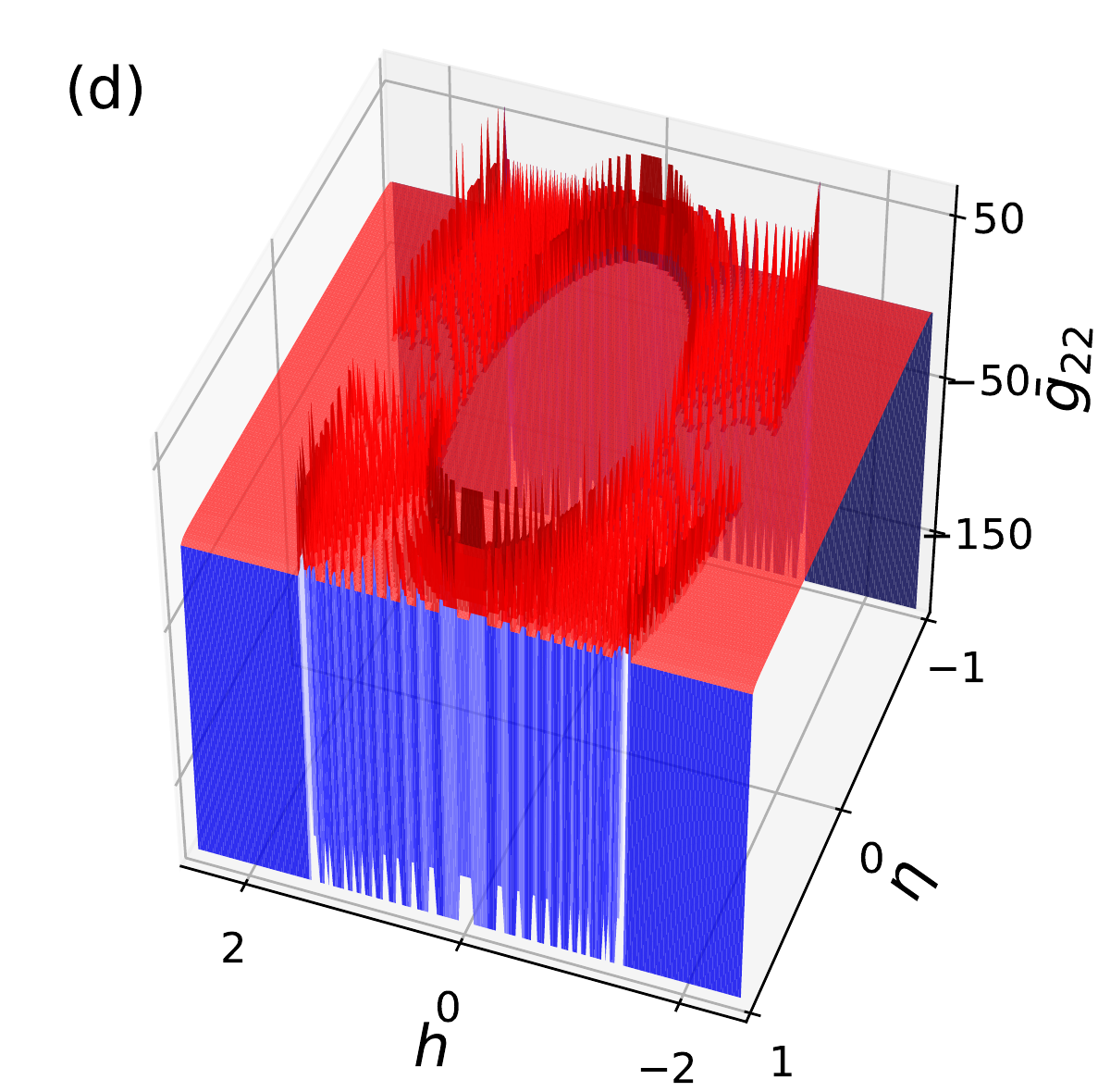}\label{fig4}}
  \caption{Intensity of the metric tensor vs system parameters, in the unbroken regime of a dimerized $XY$ model in an alternating complex magnetic field. See Supplementary Material \cite{1QGT_SM} for details of this model.  (a) $\bar{g}_{11}$ and (b) $\bar{g}_{22}$ for the anisotropic case and (c) $\bar{g}_{11}$ and (d) $\bar{g}_{22}$ for the pseudo-isotropic case. Parameters used are $J=1$, $J_s=1/2$, $\Gamma=1/3$, $\Gamma_s=1/6$ for (a) and (b), and $J=1$, $J_s=1/2$, $\Gamma=1/4$, $\Gamma_s=1/2$ for (c) and (d).
  }
  \label{figmain}
\end{figure}

Let us now furnish a concrete example to demonstrate the usefulness of the metric tensor in detecting critical points. Consider the dimerized $XY$ model in an alternating complex magnetic field \cite{1975Perk319,2010Giorgi52404}. Its Hamiltonian  reads
$H=\sum_{l=1}^L\frac{J+(-1)^lJ_s}{2}(\sigma_l^x
\sigma_{l+1}^x+\sigma_l^y
\sigma_{l+1}^y)+\frac{\Gamma+(-1)^l\Gamma_s}{2}(\sigma_l^x
\sigma_{l+1}^x-\sigma_l^y
\sigma_{l+1}^y)-\frac{h-i(-1)^l\eta}{2}\sigma_l^z$. Here, $L$ is even and $\sigma_l^\alpha$, $\alpha=x,y,z$, denote the Pauli matrices. $J>0$ and $\Gamma>0$ are homogeneous parts of coupling strengths, and $J_s>0$ and $\Gamma_s>0$ describe the amounts of staggering. The last term stands for the complex magnetic field with its strength described by $h$ and $\eta$. Since our aim is to study the role of the magnetic field in quantum criticality, we treat $h$ and $\eta$ as $\lambda$, i.e., $\lambda^{1,2}=h,\eta$, and $J,J_s,\Gamma,\Gamma_s$ as given constants. In Supplemental
Material \cite{1QGT_SM}, we analytically work out the critical points. The model
exhibits different critical behaviors for the anisotropic case, i.e., $J\Gamma\neq J_s\Gamma_s$, and the so-called pseudo-isotropic case, i.e., $J\Gamma=J_s\Gamma_s$. To simplify our analysis, for the anisotropic case, we focus on the situation $J\Gamma_s=J_s\Gamma$. There are two critical fields, i.e., $r_{c,1}=2\sqrt{J^2-\Gamma_s^2}$ and $r_{c,2}=2\sqrt{J_s^2-\Gamma^2}$, which determine the QPT points. Here, $r:=\sqrt{h^2+\eta^2}$ denotes the modulus of the field strength. For the anisotropic case, the QPT points are the $(h,\eta)$'s with $r=r_{c,1}, r_{c,2}$, whereas for the pseudo-isotropic case, they are the $(h,\eta)$'s with $r$ lying between $r_{c,1}$ and $r_{c,2}$, i.e., $\min\{r_{c,1},r_{c,2}\}\leq r\leq\max\{r_{c,1},r_{c,2}\}$. For both cases, $\mathcal{PT}$ symmetry is unbroken if and only if $\abs{\eta}<\eta_c$, where $\eta_c:=\min\{2J,2J_s\}$. So the $\mathcal{PT}$PT points are the $(h,\eta)$'s with $\eta=\eta_c$. To substantiate the usefulness of the metric tensor, we numerically compute its intensity, i.e., $\bar{g}_{\mu\nu}:=\lim_{L\rightarrow\infty}g_{0,\mu\nu}/L$, which is of interest in the thermodynamic limit $L\rightarrow\infty$ \cite{1QGT_SM}. The numerical results are presented in Fig. \ref{figmain}. For the anisotropic case, $\bar{g}_{11}$ gets very large at $r=r_{c,1}, r_{c,2}$, i.e., the QPT points, as can be seen in Fig.~\ref{fig1}. Moreover, Fig.~\ref{fig2} shows that $\bar{g}_{22}$ diverges when $\abs{\eta}$ approaches $\eta_c$, i.e., the $\mathcal{PT}$PT points. Plots of $\bar{g}_{01}$ and $\bar{g}_{10}$ are not shown here, as they do not provide additional information. For the pseudo-isotropic case shown in Figs. \ref{fig3} and \ref{fig4}, $\bar{g}_{11}$ is seen to display singular behavior so long as $r$ lies between $r_{c,1}$ and $r_{c,2}$ (consistent with our theoretical QPT analysis), and $\bar{g}_{22}$ diverges when $\abs{\eta}$ approaches $\eta_c$. Indeed, all the locations of the singularities of $\bar{g}_{\mu\nu}$ agree with the critical points found analytically \cite{1QGT_SM}, thus confirming the usefulness of the metric tensor.

In passing, with this successful extension of the geometric approach to QPTs~\cite{2007Zanardi100603} to $\mathcal{PT}$-symmetric systems,
many other geometric methods, such as those based on geometric phases \cite{2005Carollo157203,2006Zhu77206}, fidelity \cite{2006Zanardi31123,2007Cozzini14439,
2007Cozzini104420,2007Zanardi32109}, and the seminal QGT \cite{2007Venuti95701}, may be all generalized to $\mathcal{PT}$-symmetric systems.

Before concluding, we point out that $\ud s^2$ may be Riemannian or pseudo-Riemannian, depending on which variance appearing in Eq.~(\ref{eq:metric-tensor-variance}) is dominant. The pseudo-Riemannian feature of $\ud s^2$ has no counterpart in standard QM. The pseudo-Riemannian metric $\ud s^2$ is analogous to the Minkowski metric in special relativity.
We may then classify the evolution with $\ud s^2>0$, $\ud s^2=0$, and $\ud s^2<0$ as spacelike, lightlike, and timelike, respectively.  Besides,
in an accompanying paper \cite{2018Zhang-accompanying}, we extend our present results in several aspects and further show that
they admit differential-geometry interpretations.

In conclusion, we have presented an extended QGT in $\mathcal{PT}$QM.
It gives a neat, unified picture depicting the geometry of $\mathcal{PT}$QM, naturally yielding a series of geometric concepts, namely, the Berry phase (\ref{cp:BP}), the Berry curvature (\ref{cp:BCU-component}), the fidelity (\ref{df:fidelity}), and the metric tensor (\ref{cp:metric-tensor}), most of which are also formulated in this Letter. As an illustration of the usefulness of our results, we have shown how the metric tensor can be used to detect quantum criticality of $\mathcal{PT}$-symmetric systems. We believe that the extended QGT advocated  here will be highly useful in understanding and exploring more aspects of  $\mathcal{PT}$-symmetric systems.

\begin{acknowledgments}
J.G.~is supported by Singapore Ministry of Education Academic
Research Fund Tier I (WBS No.~R-144-000-353-112) and by the
Singapore NRF grant No. NRFNRFI2017-04 (WBS No. R-144-000-378-281).
Q.W.~is supported by Singapore Ministry of Education Academic
Research Fund Tier I (WBS No.~R-144-000-352-112).
D.-J.~Z.~acknowledges support from the National Natural Science Foundation of
China through Grant No.~11705105 before he joined NUS.
\end{acknowledgments}

%

\onecolumngrid
\clearpage

\renewcommand{\theequation}{\thesubsection S.\arabic{equation}}
\setcounter{equation}{0}

\section*{\large{Supplemental Material}}

\section{derivation of Eq.~(9)}

In this section, we present a derivation of Eq.~(9) in the main text. Expanding $\bra{\Phi_n(\lambda+\delta\lambda)}$ up to second order gives
\begin{eqnarray}\label{eq:expansion1}
\bra{\Phi_n(\lambda+\delta\lambda)}=
\bra{\Phi_n(\lambda)}+\bra{\partial_{\mu}\Phi_n(\lambda)}
\ud\lambda^\mu+\frac{1}{2}\bra{\partial_{\mu}
\partial_{\nu}\Phi_n(\lambda)}
\ud\lambda^\mu\ud\lambda^\nu.
\end{eqnarray}
Contracting both sides of Eq.~(\ref{eq:expansion1}) with $\ket{\Psi_n(\lambda)}$, we have
\begin{eqnarray}\label{eq:expansion1-1}
\inp{\Phi_n(\lambda+\delta\lambda)}{\Psi_n(\lambda)}=
1+\inp{\partial_{\mu}\Phi_n(\lambda)}{\Psi_n(\lambda)}
\ud\lambda^\mu+\frac{1}{2}\inp{\partial_{\mu}
\partial_{\nu}\Phi_n(\lambda)}{\Psi_n(\lambda)}
\ud\lambda^\mu\ud\lambda^\nu.
\end{eqnarray}
Similarly, expanding $\ket{\Psi_n(\lambda+\delta\lambda)}$ up to second order and contracting its both sides with $\bra{\Phi_n(\lambda)}$, we obtain
\begin{eqnarray}\label{eq:expansion2}
\inp{\Phi_n(\lambda)}{\Psi_n(\lambda+\delta\lambda)}=
1+\inp{\Phi_n(\lambda)}{\partial_{\mu}\Psi_n(\lambda)}
\ud\lambda^\mu+\frac{1}{2}\inp{\Phi_n(\lambda)}{\partial_{\mu}
\partial_{\nu}\Psi_n(\lambda)}
\ud\lambda^\mu\ud\lambda^\nu.
\end{eqnarray}
From Eqs. (\ref{eq:expansion1-1}) and (\ref{eq:expansion2}), it follows that the product of $\inp{\Phi_n(\lambda+\delta\lambda)}{\Psi_n(\lambda)}$ and
$\inp{\Phi_n(\lambda)}{\Psi_n(\lambda+\delta\lambda)}$ reads, up to second order,
\begin{eqnarray}\label{eq:1}
&&\inp{\Phi_n(\lambda+\delta\lambda)}{\Psi_n(\lambda)}
\inp{\Phi_n(\lambda)}{\Psi_n(\lambda+\delta\lambda)}
\nonumber\\
&&=1+\left(\inp{\partial_{\mu}\Phi_n(\lambda)}{\Psi_n(\lambda)}
\inp{\Phi_n(\lambda)}{\partial_{\nu}\Psi_n(\lambda)}
+\frac{1}{2}\inp{\partial_{\mu}
\partial_{\nu}\Phi_n(\lambda)}{\Psi_n(\lambda)}
+\frac{1}{2}\inp{\Phi_n(\lambda)}{\partial_{\mu}
\partial_{\nu}\Psi_n(\lambda)}\right)
\ud\lambda^\mu\ud\lambda^\nu.
\end{eqnarray}
On the other hand, note that $\inp{\partial_{\nu}\Phi_n(\lambda)}{\Psi_n(\lambda)}
+\inp{\Phi_n(\lambda)}{\partial_{\nu}\Psi_n(\lambda)}=0$.
Differentiating both sides of this equality w.r.t.~$\lambda^\mu$ gives
\begin{eqnarray}\label{eq:2}
\inp{\partial_{\mu}
\partial_{\nu}\Phi_n(\lambda)}{\Psi_n(\lambda)}
+\inp{\Phi_n(\lambda)}{\partial_{\mu}
\partial_{\nu}\Psi_n(\lambda)}
=-\inp{\partial_{\mu}\Phi_n(\lambda)}{\partial_{\nu}
\Psi_n(\lambda)}-\inp{\partial_{\nu}\Phi_n(\lambda)}
{\partial_{\mu}
\Psi_n(\lambda)}.
\end{eqnarray}
Substituting Eq.~(\ref{eq:2}) into Eq.~(\ref{eq:1}),
we have
\begin{eqnarray}\label{eq:3}
&&\inp{\Phi_n(\lambda+\delta\lambda)}{\Psi_n(\lambda)}
\inp{\Phi_n(\lambda)}{\Psi_n(\lambda+\delta\lambda)}
=\nonumber\\
&&1-\frac{1}{2}\left(\inp{\partial_{\mu}
\Phi_n(\lambda)}{\partial_{\nu}
\Psi_n(\lambda)}+\inp{\partial_{\nu}\Phi_n(\lambda)}
{\partial_{\mu}
\Psi_n(\lambda)}-2
\inp{\partial_{\mu}\Phi_n(\lambda)}{\Psi_n(\lambda)}
\inp{\Phi_n(\lambda)}{\partial_{\nu}\Psi_n(\lambda)}\right)
\ud\lambda^\mu\ud\lambda^\nu.
\end{eqnarray}
Inserting Eq. (\ref{eq:3}) into the expression $\ud s^2=2(1-\sqrt{\abs{\inp{\Phi_n(\lambda+\delta\lambda)}{\Psi_n(\lambda)}
\inp{\Phi_n(\lambda)}{\Psi_n(\lambda+\delta\lambda)}
}})$ and noting that $\sqrt{\abs{1-z}}=\frac{1}{2}\Re(z)$ for a small complex number $z$, we obtain
\begin{eqnarray}\label{eq:4}
\ud s^2=\frac{1}{2}\Re\left(\inp{\partial_{\mu}\Phi_n(\lambda)}
{\partial_{\nu}
\Psi_n(\lambda)}+\inp{\partial_{\nu}\Phi_n(\lambda)}
{\partial_{\mu}
\Psi_n(\lambda)}-2
\inp{\partial_{\mu}\Phi_n(\lambda)}{\Psi_n(\lambda)}
\inp{\Phi_n(\lambda)}{\partial_{\nu}\Psi_n(\lambda)}\right)
\ud\lambda^\mu\ud\lambda^\nu.
\end{eqnarray}
Since $\Re\inp{\partial_{\nu}\Phi_n(\lambda)}
{\partial_{\mu}
\Psi_n(\lambda)}=\Re\inp{\partial_{\mu}
\Psi_n(\lambda)}{\partial_{\nu}\Phi_n(\lambda)}$ and
$\Re\inp{\partial_{\mu}\Phi_n(\lambda)}{\Psi_n(\lambda)}
\inp{\Phi_n(\lambda)}{\partial_{\nu}\Psi_n(\lambda)}
=\Re\inp{\partial_{\mu}\Psi_n(\lambda)}{\Phi_n(\lambda)}
\inp{\Psi_n(\lambda)}{\partial_{\nu}\Phi_n(\lambda)}$,
we can rewrite Eq. (\ref{eq:4}) as
\begin{eqnarray}
\ud s^2=g_{n,\mu\nu}\ud\lambda^\mu\ud\lambda^\nu,
\end{eqnarray}
with
\begin{eqnarray}\label{eq:5}
g_{n,\mu\nu}
=&&\frac{1}{2}\Re(\inp{\partial_{\mu}\Phi_n(\lambda)}
{\partial_{\nu}
\Psi_n(\lambda)}-
\inp{\partial_{\mu}\Phi_n(\lambda)}{\Psi_n(\lambda)}
\inp{\Phi_n(\lambda)}{\partial_{\nu}\Psi_n(\lambda)}
\nonumber\\
&&+
\inp{\partial_{\mu}\Psi_n(\lambda)}{\partial_{\nu}
\Phi_n(\lambda)}-
\inp{\partial_{\mu}\Psi_n(\lambda)}{\Phi_n(\lambda)}
\inp{\Psi_n(\lambda)}{\partial_{\nu}\Phi_n(\lambda)}).
\end{eqnarray}
Equation (\ref{eq:5}) is exactly the metric tensor expressed by Eq. (9) in the main text.

\section{Derivation of the invariance of Eq.~(10)}

In this section, we show that Eq.~(10) in the main text is independent of any specific choices of $W(\lambda)$, as long as $W(\lambda)$ satisfies $W(\lambda)H(\lambda)=H^\dagger(\lambda)W(\lambda)$. Note that $\ket{\Psi_n(\lambda)}$ is the $n$-th eigenstate of $H(\lambda)$ by definition. If a different $W(\lambda)$ is chosen, $\ket{\Psi_n(\lambda)}$ must transform as
\begin{eqnarray}
\ket{\Psi_n(\lambda)}\rightarrow\ket{\widetilde{\Psi}_n(\lambda)}:=f(\lambda)\ket{\Psi_n(\lambda)},
\end{eqnarray}
where $f(\lambda)$ is a certain complex-valued function. On the other hand, since
\begin{eqnarray}
H^\dagger(\lambda)\ket{\Phi_n(\lambda)}=H^\dagger(\lambda)
W(\lambda)\ket{\Psi_n(\lambda)}=W(\lambda)H(\lambda)
\ket{\Psi_n(\lambda)}=E_n(\lambda)W(\lambda)\ket{\Psi_n(\lambda)}
=E_n(\lambda)\ket{\Phi_n(\lambda)},
\end{eqnarray}
$\ket{\Phi_n(\lambda)}$ is the $n$-th eigenstate of $H^\dagger(\lambda)$. So, similar to $\ket{\Psi_n(\lambda)}$, $\ket{\Phi_n(\lambda)}$ must transform as
\begin{eqnarray}
\ket{\Phi_n(\lambda)}\rightarrow
\ket{\widetilde{\Phi}_n(\lambda)}=g(\lambda)\ket{\Phi_n(\lambda)},
\end{eqnarray}
where $g(\lambda)$ is another complex-valued function. Besides, the normalization condition $\inpp{\widetilde{\Psi}_n(\lambda)}
{\widetilde{\Psi}_n(\lambda)}_\lambda=1$,
i.e., $\inp{\widetilde{\Phi}_n(\lambda)}{\widetilde{\Psi}_n(\lambda)}=1$,
requires that
\begin{eqnarray}
f(\lambda)g^*(\lambda)=1.
\end{eqnarray}
Now, it is straightforward to verify the invariance of Eq.~(10) in the main text by inserting $\ket{\widetilde{\Psi}_n(\lambda)}=f(\lambda)\ket{\Psi_n(\lambda)}$,
$\ket{\widetilde{\Phi}_n(\lambda)}=g(\lambda)\ket{\Phi_n(\lambda)}$, and $f(\lambda)g^*(\lambda)=1$
into it.

\section{derivation of Eq.~(13)}

In analogy to the conventional Hermitian operator, the eigenstates of the physical Hermitian operator $H(\lambda)$, i.e., $\ket{\Psi_n(\lambda)}$, forms a complete set and satisfy $\inpp{\Psi_m(\lambda)}{\Psi_n(\lambda)}_\lambda=\delta_{mn}$. This amounts to the fact
\begin{eqnarray}
\inp{\Phi_m(\lambda)}{\Psi_n(\lambda)}=\delta_{mn},
~~~~
\sum_{n}\ket{\Psi_n(\lambda)}\bra{\Phi_n(\lambda)}=I.
\end{eqnarray}
That is, $\ket{\Psi_n(\lambda)}$ and $\ket{\Phi_n(\lambda)}$ constitute a biorthonormal basis.
Using the equality $\sum_{n}\ket{\Psi_n(\lambda)}\bra{\Phi_n(\lambda)}=I$, we can rewrite Eq.~(9) in the main text as
\begin{eqnarray}\label{GS-Metric}
g_{0,\mu\nu}=\frac{1}{2}\Re\left(\sum_{n\neq 0}\inp{\partial_\mu\Phi_0(\lambda)}{\Psi_n(\lambda)}
\inp{\Phi_n(\lambda)}{\partial_\nu\Psi_0(\lambda)}
+\inp{\partial_\mu\Psi_0(\lambda)}{\Phi_n(\lambda)}
\inp{\Psi_n(\lambda)}{\partial_\nu\Phi_0(\lambda)}\right).
\end{eqnarray}
On the other hand, in terms of this biorthonormal basis, $H(\lambda)$ can be expressed as
\begin{eqnarray}\label{H-lam}
H(\lambda)=\sum_nE_n(\lambda)\ket{\Psi_n(\lambda)}\bra{\Phi_n(\lambda)}.
\end{eqnarray}
This point can be easily verified by noting that $H(\lambda)$ in Eq.~(\ref{H-lam}) satisfies $H(\lambda)\ket{\Psi_n(\lambda)}=E_n(\lambda)\ket{\Psi_n(\lambda)}$.
From Eq.~(\ref{H-lam}), it follows that
\begin{eqnarray}
\inp{\partial_\mu\Phi_m(\lambda)}{\Psi_n(\lambda)}=
\frac{\bra{\Phi_m(\lambda)}\partial_\mu H(\lambda)\ket{\Psi_n(\lambda)}}{E_m(\lambda)-E_n(\lambda)},
\end{eqnarray}
provided that $E_m(\lambda)\neq E_n(\lambda)$. Substituting this equation into Eq.~(\ref{GS-Metric}), we obtain, after some algebra,
\begin{eqnarray}
g_{0,\mu\nu}=\Re\sum_{n\neq 0}
\frac{\bra{\Phi_0(\lambda)}\partial_\mu H(\lambda)\ket{\Psi_n(\lambda)}\bra{\Phi_n(\lambda)}
\partial_\nu H(\lambda)\ket{\Psi_0(\lambda)}+\bra{\Phi_n(\lambda)}
\partial_\mu
H(\lambda)\ket{\Psi_0(\lambda)}\bra{\Phi_0(\lambda)}
\partial_\nu
H(\lambda)\ket{\Psi_n(\lambda)}}
{2\left[E_0(\lambda)-E_n(\lambda)\right]^2},\nonumber\\
\end{eqnarray}
which is exactly Eq.~(13) in the main text.

\section{Derivation of the expression of $O_{B,\mu}$}

In this section, we present a derivation of the expression of $O_{B,\mu}$, i.e., $O_{B,\mu}=-W(\lambda)^{-1}\partial_\mu W(\lambda)/2$, in the main text. Using the closure relation $\sum_n\ket{\Psi_n(\lambda)}
\bra{\Phi_n(\lambda)}=I$, we can rewrite the defining expression of $O_\mu$, i.e.,  $O_\mu=i\sum_n\ket{\partial_\mu\Psi_n(\lambda)}
\bra{\Phi_n(\lambda)}$, as follows:
\begin{eqnarray}\label{eq:O}
O_\mu=-i\sum_n\ket{\Psi_n(\lambda)}
\bra{\partial_\mu\Phi_n(\lambda)}.
\end{eqnarray}
From Eq.~(\ref{eq:O}) and noting that $W(\lambda)=\sum_n\ket{\Phi_n(\lambda)}\bra{\Phi_n(\lambda)}$, we deduce that
\begin{eqnarray}\label{eq:WO}
W(\lambda) O_\mu=-i\sum_n\ket{\Phi_n(\lambda)}
\bra{\partial_\mu\Phi_n(\lambda)},
\end{eqnarray}
and
\begin{eqnarray}\label{eq:OW}
O_\mu^\dagger W(\lambda) =i\sum_n\ket{\partial_\mu\Phi_n(\lambda)}
\bra{\Phi_n(\lambda)}.
\end{eqnarray}
From Eqs. (\ref{eq:WO}) and (\ref{eq:OW}), it follows immediately that
\begin{eqnarray}\label{eq:WO-OW}
W(\lambda) O_\mu-O_\mu^\dagger W(\lambda)=-i\partial_\mu W(\lambda).
\end{eqnarray}
On the other hand, note that an operator $X$ is Hermitian w.r.t. the $\lambda$-dependent inner product if and only if it satisfies $W(\lambda) X=X^\dagger W(\lambda)$. Decomposing $O_\mu$ as $O_\mu=O_{A,\mu}+iO_{B,\mu}$, with $O_{A,\mu}$ and $O_{B,\mu}$ being two physical Hermitian operators, we have
$W(\lambda) O_\mu=W(\lambda) O_{A,\mu}+iW(\lambda) O_{B,\mu}$ and
$O_\mu^\dagger W(\lambda) =W(\lambda) O_{A,\mu}-iW(\lambda) O_{B,\mu}$. Substituting these two equalities into Eq.~(\ref{eq:WO-OW}) yields
\begin{eqnarray}
O_{B,\mu}=-\frac{1}{2}W(\lambda)^{-1}\partial_\mu W(\lambda).
\end{eqnarray}
This completes the derivation of the expression of $O_{B,\mu}$ in the main text.

\section{Derivation of Eq.~(14)}
In this section, we present a derivation of Eq.~(14) in the main text. As stated in the main text, $O_\mu$ satisfies the equations
\begin{eqnarray}\label{eq:Sch1}
i\ket{\partial_\mu\Psi_n(\lambda)}=O_\mu\ket{\Psi_n(\lambda)}
\end{eqnarray}
and
\begin{eqnarray}\label{eq:Sch2}
i\ket{\partial_\mu\Phi_n(\lambda)}=O_\mu^\dagger
\ket{\Phi_n(\lambda)}.
\end{eqnarray}
Substituting Eqs.~(\ref{eq:Sch1}) and (\ref{eq:Sch2}) into Eq.~(9) in the main text gives
\begin{eqnarray}\label{eq:g}
g_{0,\mu\nu}&=&\frac{1}{2}\Re(\bra{\Phi_0(\lambda)}O_\mu O_\nu\ket{\Psi_0(\lambda)}-\bra{\Phi_0(\lambda)}O_\mu\ket{\Psi_0
(\lambda)}
\bra{\Phi_0(\lambda)}O_\nu\ket{\Psi_0(\lambda)}
\nonumber\\
&+&\bra{\Psi_0(\lambda)}O_\mu^\dagger O_\nu^\dagger\ket{\Phi_0(\lambda)}-\bra{\Psi_0(\lambda)}O_\mu^\dagger
\ket{\Phi_0(\lambda)}
\bra{\Psi_0(\lambda)}O_\nu^\dagger\ket{\Phi_0(\lambda)}).
\end{eqnarray}
Using the decomposition $O_\mu=O_{A,\mu}+iO_{B,\nu}$ and noting that $\ket{\Phi_0(\lambda)}=W(\lambda)\ket{\Psi_0(\lambda)}$, we have, after some simple algebra,
\begin{eqnarray}\label{eq:g1}
\bra{\Phi_0(\lambda)}O_\mu O_\nu\ket{\Psi_0(\lambda)}
=\exppt{O_{A,\mu}O_{A,\nu}}_\lambda-
\exppt{O_{B,\mu}O_{B,\nu}}_\lambda
+i\exppt{O_{A,\mu}O_{B,\nu}}_\lambda
+i\exppt{O_{B,\mu}O_{A,\nu}}_\lambda,
\end{eqnarray}
\begin{eqnarray}\label{eq:g2}
&&\bra{\Phi_0(\lambda)}O_\mu\ket{\Psi_0(\lambda)}
\bra{\Phi_0(\lambda)}O_\nu\ket{\Psi_0(\lambda)}
\nonumber\\
&&=\exppt{O_{A,\mu}}_\lambda\exppt{O_{A,\nu}}_\lambda-
\exppt{O_{B,\mu}}_\lambda\exppt{O_{B,\nu}}_\lambda
+i\exppt{O_{A,\mu}}_\lambda\exppt{O_{B,\nu}}_\lambda
+i\exppt{O_{B,\mu}}_\lambda\exppt{O_{A,\nu}}_\lambda,
\end{eqnarray}
\begin{eqnarray}\label{eq:g3}
\bra{\Psi_0(\lambda)}O_\mu^\dagger O_\nu^\dagger\ket{\Phi_0(\lambda)}=
\exppt{O_{A,\mu}O_{A,\nu}}_\lambda-
\exppt{O_{B,\mu}O_{B,\nu}}_\lambda
-i\exppt{O_{A,\mu}O_{B,\nu}}_\lambda
-i\exppt{O_{B,\mu}O_{A,\nu}}_\lambda,
\end{eqnarray}
\begin{eqnarray}\label{eq:g4}
&&\bra{\Psi_0(\lambda)}O_\mu^\dagger
\ket{\Phi_0(\lambda)}
\bra{\Psi_0(\lambda)}O_\nu^\dagger\ket{\Phi_0(\lambda)}
\nonumber\\
&&=\exppt{O_{A,\mu}}_\lambda\exppt{O_{A,\nu}}_\lambda-
\exppt{O_{B,\mu}}_\lambda\exppt{O_{B,\nu}}_\lambda
-i\exppt{O_{A,\mu}}_\lambda\exppt{O_{B,\nu}}_\lambda
-i\exppt{O_{B,\mu}}_\lambda\exppt{O_{A,\nu}}_\lambda.
\end{eqnarray}
Inserting Eqs. (\ref{eq:g1}), (\ref{eq:g2}), (\ref{eq:g3}), and (\ref{eq:g4}) into Eq. (\ref{eq:g}), we have
\begin{eqnarray}\label{eq:g5}
g_{0,\mu\nu}=\Re\left(\exppt{O_{A,\mu}O_{A,\nu}}_\lambda-
\exppt{O_{A,\mu}}_\lambda\exppt{O_{A,\nu}}_\lambda
-\exppt{O_{B,\mu}O_{B,\nu}}_\lambda+
\exppt{O_{B,\mu}}_\lambda\exppt{O_{B,\nu}}_\lambda\right).
\end{eqnarray}
Note that
\begin{eqnarray}\label{eq:g6}
\Re\exppt{O_{A,\mu}O_{A,\nu}}_\lambda=
\frac{1}{2}\Re(\exppt{O_{A,\mu}O_{A,\nu}}_\lambda+
\exppt{O_{A,\nu}O_{A,\mu}}_\lambda),
\end{eqnarray}
\begin{eqnarray}\label{eq:g7}
\Re\exppt{O_{B,\mu}O_{B,\nu}}_\lambda=
\frac{1}{2}\Re(\exppt{O_{B,\mu}O_{B,\nu}}_\lambda+
\exppt{O_{B,\nu}O_{B,\mu}}_\lambda).
\end{eqnarray}
Using Eqs. (\ref{eq:g6}) and (\ref{eq:g7})
and letting $\bar{O}_{A,\mu}=O_{A,\mu}-\exppt{O_{A,\mu}}_\lambda$
and $\bar{O}_{B,\mu}=O_{B,\mu}-\exppt{O_{B,\mu}}_\lambda$,
we can rewrite Eq. (\ref{eq:g5}) as
\begin{eqnarray}
g_{0,\mu\nu}=\frac{1}{2}
\left(\exppt{\{\bar{O}_{A,\mu},\bar{O}_{A,\nu}\}}_\lambda
-\exppt{\{\bar{O}_{B,\mu},
\bar{O}_{B,\nu}\}}_\lambda\right),
\end{eqnarray}
which is exactly Eq. (14) in the main text.

\section{Details of the example}

In this section, we present details of calculations regarding the example in the main text. The model considered in the example is the dimerized $XY$ model in an alternating complex magnetic field \cite{1975Perk319b,2010Giorgi52404b}. Its Hamiltonian reads
\begin{eqnarray}\label{H}
H=\sum_{l=1}^L\frac{J+(-1)^lJ_s}{2}(\sigma_l^x
\sigma_{l+1}^x+\sigma_l^y
\sigma_{l+1}^y)+\frac{\Gamma+(-1)^l\Gamma_s}{2}(\sigma_l^x
\sigma_{l+1}^x-\sigma_l^y
\sigma_{l+1}^y)-\frac{h-i(-1)^l\eta}{2}\sigma_l^z.
\end{eqnarray}
Here, $L$ is an even integer representing the number of spins, and $\sigma_l^\alpha$, $\alpha=x,y,z$, denote the Pauli matrices at the site $l$. $\sigma_{L+1}^\alpha$ is understood as $\sigma_{1}^\alpha$, i.e., the periodic boundary condition is assumed. The first term of Eq. (\ref{H}) stands for the isotropic interactions between neighbouring spins. $J>0$ represents the strength of the homogeneous part and $J_s>0$ is the amount of staggering, i.e., the interaction strengths for even and odd values of $l$ are $J+J_s$ and $J-J_s$, respectively. $\Gamma>0$ and $\Gamma_s>0$ describe the normal and staggered anisotropic interactions. The last term in Eq. (\ref{H}) is the influence of an alternating external magnetic field with the complex strength described by $h$ and $\eta$. The alternating nature of the field is assumed to arise from different magnetic moments of the spins at even and odd lattice sites, respectively. In the special case of $J_s=\Gamma_s=\eta=0$, the Hamiltonian (\ref{H}) reduces to the Hamiltonian of the usual $XY$ model with the anisotropy parameter $\gamma=\Gamma J^{-1}$. On the other hand, the Hamiltonian (\ref{H}) is $\mathcal{PT}$-symmetric. Indeed, for this model, the effects of $\mathcal{T}$ and $\mathcal{P}$ read $\mathcal{T}i\mathcal{T}=-i$ and $\mathcal{P}\sigma_l^\alpha\mathcal{P}=\sigma_{L+1-l}^\alpha$
\cite{2010Giorgi52404}.
It is easy to verify that $[\mathcal{P},H]\neq 0$ and $[\mathcal{T},H]\neq 0$ but $[\mathcal{PT},H]=0$. In the following, we aim to study the role of the complex magnetic field in quantum criticality, thus treating parameters $h$ and $\eta$ as $\lambda$, i.e., $\lambda^1=h$ and $\lambda^2=\eta$, and $J$, $J_s$, $\Gamma$, $\Gamma_s$ as given constants. As for quantum criticality in the model, one need to distinguish two cases, the anisotropic case, i.e., $J\Gamma\neq J_s\Gamma_s$, and the pseudo-isotropic case, i.e., $J\Gamma=J_s\Gamma_s$. The two cases belong to different universality classes. To simplify our discussion, for the anisotropic case, we restrict ourselves to the situation $J\Gamma_s=J_s\Gamma$. In the following, we identify the ground-state phase diagram of the model, as presented in Fig. \ref{fig}, step by step.
\begin{figure}
  \centering
  \includegraphics[width=.4\textwidth]{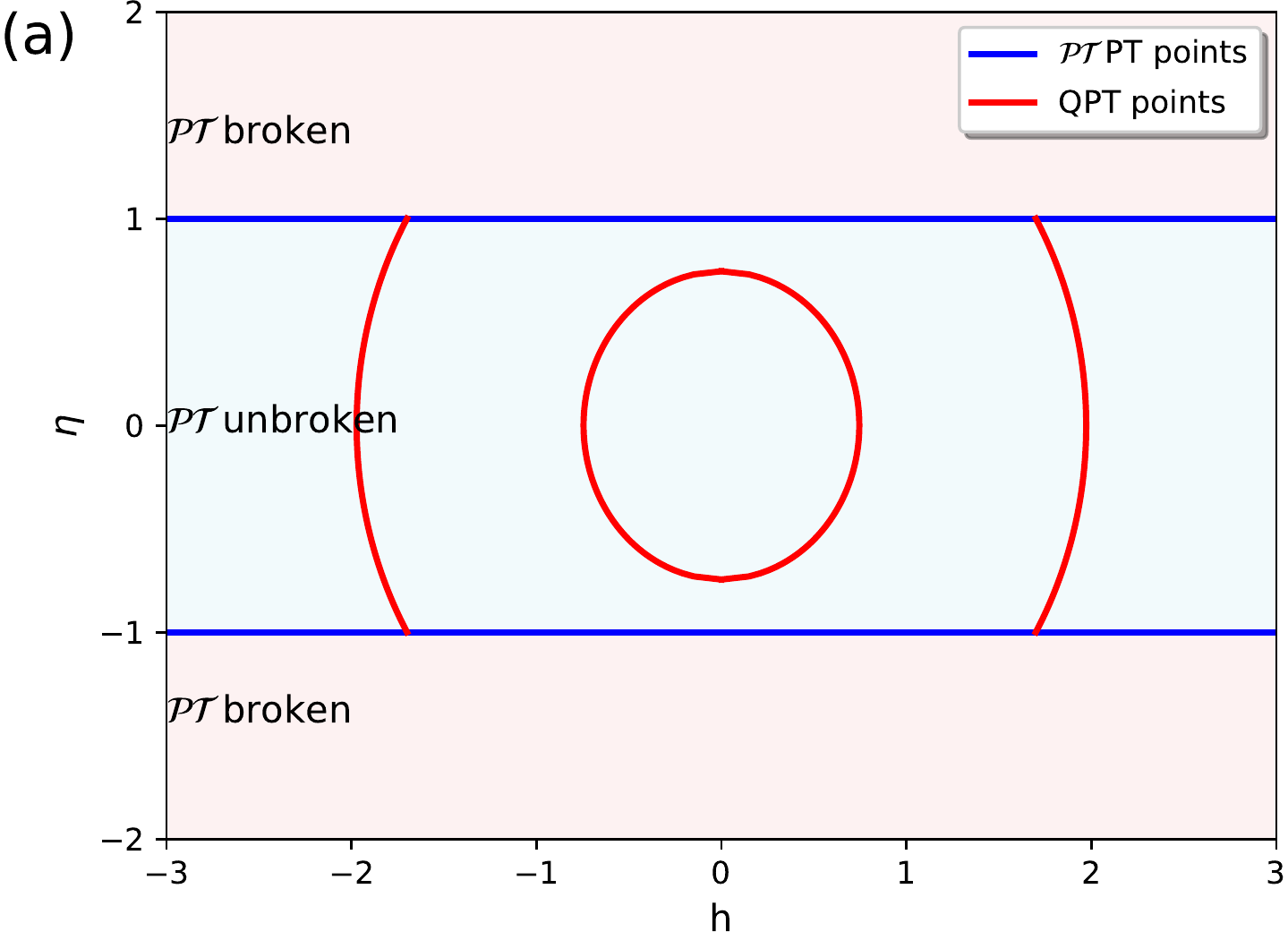}\hspace{.2in}
  \includegraphics[width=.4\textwidth]{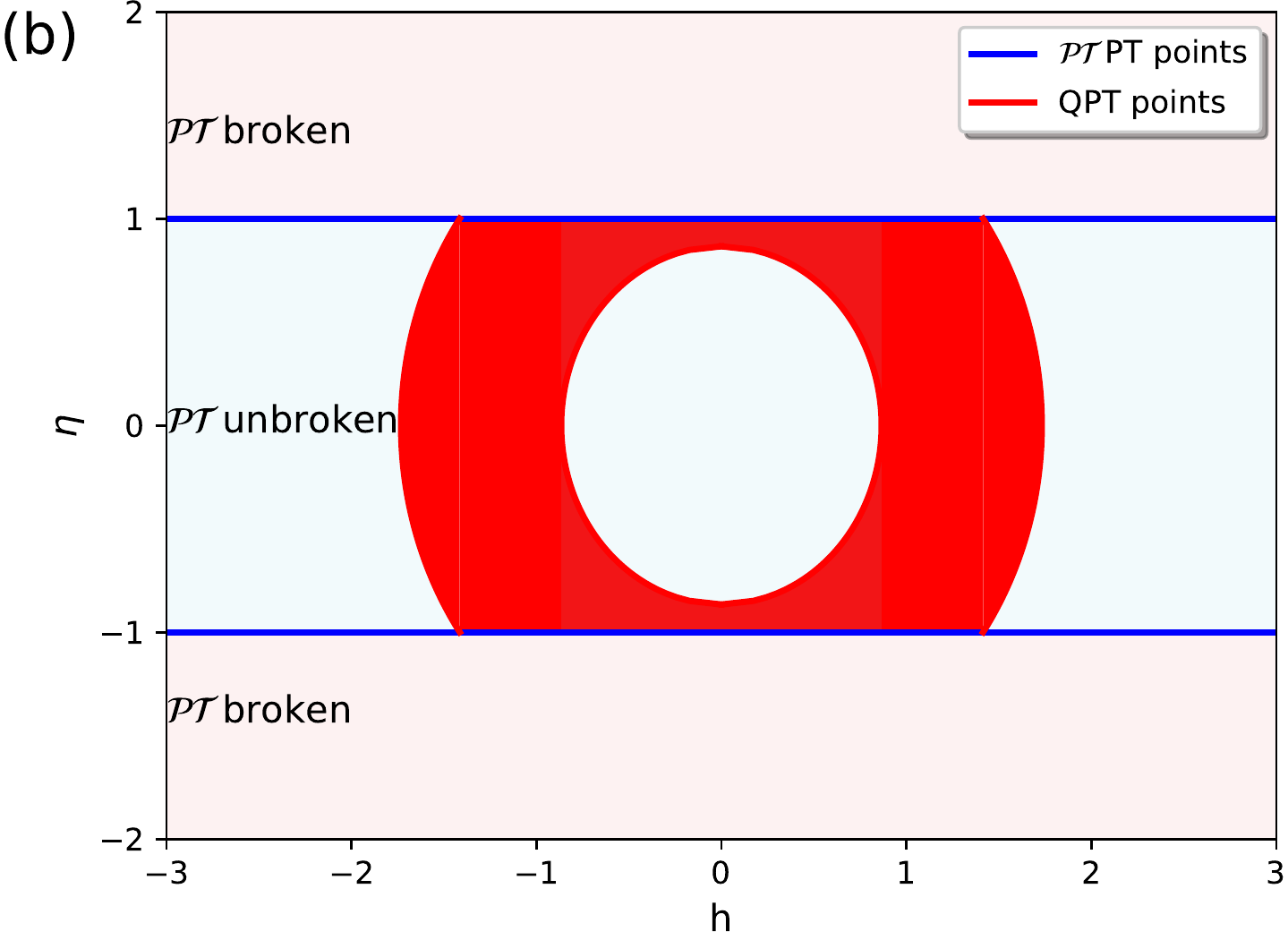}
  \caption{Ground-state phase diagram of the model: (a) for the anisotropic case and (b) for the pseudo-isotropic case. The $\mathcal{PT}$ unbroken regime and broken regime are separated by the $\mathcal{PT}$PT points (constituting the blue curves) for both (a) and (b). The QPT points are in red color. Parameters used are $J=1$, $J_s=1/2$, $\Gamma=1/3$, $\Gamma_s=1/6$ for (a), and $J=1$, $J_s=1/2$, $\Gamma=1/4$, $\Gamma_s=1/2$ for (b). Accordingly, $\eta_c=1$ for both cases, $r_{c,1}=\sqrt{35}/3$ and $r_{c,2}=\sqrt{5}/3$ for the anisotropic case, and $r_{c,1}=\sqrt{3}$ and $r_{c,2}=\sqrt{3}/2$ for the pseudo-isotropic case.
  }
  \label{fig}
\end{figure}

\textit{First, we diagonalize the Hamiltonian}. The Hamiltonian (\ref{H}) can be diagonalized by a standard procedure, which can be summarized as the following three steps:

\textit{Step 1: Jordan-Wigner transformation.} In order to express the Hamiltonian in terms of fermion operators, let
\begin{eqnarray}\label{JW-transformation}
c_l:=\left(\prod_{m<l}\sigma_l^z\right)(\sigma_l^x+i\sigma_l^y)/2,
~~~~
c_l^+:=\left(\prod_{m<l}\sigma_l^z\right)(\sigma_l^x-i\sigma_l^y)/2,
\end{eqnarray}
where $c_l$ and $c_l^\dagger$ are fermion annihilation and creation operators. From Eq. (\ref{JW-transformation}), it follows that
\begin{eqnarray}\label{eq:J-W}
\sigma_l^x=\prod_{m<l}(1-2c_m^\dagger c_m)(c_l^\dagger+c_l),
~~~~
\sigma_l^y=i\prod_{m<l}(1-2c_m^\dagger c_m)(c_l^\dagger-c_l),
~~~~
\sigma_l^z=1-2c_l^\dagger c_l.
\end{eqnarray}
Substituting Eq. (\ref{eq:J-W}) into Eq. (\ref{H}) gives
\begin{eqnarray}\label{H-JW}
H=\sum_{l=1}^L\left[J+(-1)^lJ_s\right](c_l^\dagger c_{l+1}+c_{l+1}^\dagger c_l)+\left[\Gamma+(-1)^l\Gamma_s\right](c_l^\dagger c_{l+1}^\dagger+c_{l+1}c_l)+\left[h-i(-1)^l\eta\right]c_l^\dagger c_l.
\end{eqnarray}

\textit{Step 2: Fourier transformation.} Let
\begin{eqnarray}\label{F-transformation}
a_k:=\frac{1}{\sqrt{L}}\sum_{l=1}^Le^{-ikl}c_l,
\end{eqnarray}
with $-\pi<k=\frac{2\pi m}{L}\leq\pi$, where $m$ is an integer.
Using the relations
\begin{eqnarray}
&&\sum_{l=1}^L c_l^\dagger c_{l+1}=\sum_k a_k^\dagger a_k e^{ik},
~~~~
\sum_{l=1}^L(-1)^lc_l^\dagger c_{l+1}=-\sum_k a_k^\dagger a_{k-\pi} e^{ik},
~~~~
\sum_{l=1}^L c_l^\dagger c_{l+1}^\dagger=\sum_k a_k^\dagger a_{-k}^\dagger e^{ik},\\
&&\sum_{l=1}^L(-1)^lc_l^\dagger c_{l+1}^\dagger=-\sum_ka_k^\dagger a_{\pi-k}^\dagger e^{ik},
~~~~
\sum_{l=1}^L c_l^\dagger c_l=\sum_k a_k^\dagger a_k,
~~~~
\sum_{l=1}^L(-1)^l c_l^\dagger c_l=\sum_k a_k^\dagger a_{k-\pi},
\end{eqnarray}
and similar expressions for the Hermitian conjugate relations, we can write the Hamiltonian (\ref{H-JW}) in the form
\begin{eqnarray}\label{H-F}
H=\sum_{0<k<\pi/2}\bm{a}_k^\dagger\cdot D_k\cdot \bm{a}_k,
\end{eqnarray}
where $\bm{a}_k^\dagger=(a_k^\dagger,a_{-k},a_{\pi-k},a_{k-\pi}^\dagger)$
and
\begin{eqnarray}
D_k=\begin{pmatrix}
       2J\cos(k)+h & 2i\Gamma\sin(k) & -2\Gamma_s\cos(k) & -2iJ_s\sin(k)-i\eta \\
       -2i\Gamma\sin(k) & -2J\cos(k)-h & 2iJ_s\sin(k)+i\eta & 2\Gamma_s\cos(k) \\
       -2\Gamma_s\cos(k) & -2iJ_s\sin(k)+i\eta & 2J\cos(k)-h & 2i\Gamma\sin(k)\\
       2iJ_s\sin(k)-i\eta & 2\Gamma_s\cos(k) & -2i\Gamma\sin(k) & -2J\cos(k)+h
     \end{pmatrix}.
\end{eqnarray}

\textit{Step 3: generalized Bogoliubov transformation.} In the unbroken regime, to be determined later on, $D_k$ can be diagonalized using a biorthonormal bais, i.e.,
\begin{eqnarray}\label{D}
D_k=\sum_{n=1}^4\varepsilon_{n,k}\ket{\psi_{n,k}}\bra{\phi_{n,k}},
\end{eqnarray}
with $\inp{\psi_{m,k}}{\phi_{n,k}}=\delta_{mn}$ and $\sum_{n=1}^4\ket{\psi_{n,k}}\bra{\phi_{n,k}}=I_{4}$. Noting that $\{D_k,I_{2}\otimes\sigma_x\}=0$ and after tedious calculations, we find that
\begin{eqnarray}\label{Energy}
\varepsilon_{n,k}=\Lambda_{\pm}(k),~~-\Lambda_{\pm}(k),~~~~n=1,2,3,4,
\end{eqnarray}
where
\begin{eqnarray}\label{Lambda}
\Lambda_{\pm}(k):=\left[c_2(k)\pm\sqrt{c_2(k)^2-c_4(k)}
\right]^{\frac{1}{2}},
\end{eqnarray}
with
\begin{eqnarray}\label{c2-c4}
c_2(k)&=&h^2-\eta^2+4(J^2+\Gamma_s^2)\cos(k)^2+4(J_s^2+\Gamma^2)\sin(k)^2,
\\
c_4(k)&=&\left[h^2+\eta^2-4(J^2-\Gamma_s^2)\cos(k)^2-
4(J_s^2-\Gamma^2)\sin(k)^2
\right]^2+16(J\Gamma-J_s\Gamma_s)^2\sin(2k)^2.
\end{eqnarray}
The expressions of $\ket{\psi_{n,k}}$ and $\ket{\phi_{n,k}}$ are rather complicated and hence are omitted here. Inserting Eq. (\ref{D}) into Eq. (\ref{H-F}), we have that the Hamiltonian assumes the diagonal form
\begin{eqnarray}\label{H-diag}
H=\sum_{n=1}^4\sum_{0<k<\pi/2}\varepsilon_{n,k}b_{n,k}^\dagger \tilde{b}_{n,k},
\end{eqnarray}
where
\begin{eqnarray}\label{b-b}
b_{n,k}:=\bra{\psi_{n,k}}\cdot\bm{a}_k,
~~~~\tilde{b}_{n,k}:=\bra{\phi_{n,k}}\cdot\bm{a}_k.
\end{eqnarray}
It is easy to verify the following anti-commutation relations
\begin{eqnarray}\label{anti-com}
\{b_{n,k},b_{n^\prime,k^\prime}\}
=\{\tilde{b}_{n,k},\tilde{b}_{n^\prime,k^\prime}\}
=\{b_{n,k},\tilde{b}_{n^\prime,k^\prime}\}=0,
~~~~\{b_{n,k},\tilde{b}_{n^\prime,k^\prime}^\dagger\}=\delta_{nn^\prime}
\delta_{kk^\prime}.
\end{eqnarray}

\textit{Second, we identify the unbroken regime.} Intuitively, if $\eta$ is strong enough to make complex some of the eigenvalues of $H$, the $\mathcal{PT}$ symmetry is spontaneously broken.  We will show that for both cases, i.e., the anisotropic case with $J\Gamma_s=J_s\Gamma$ and the pseudo-isotropic case, the Hamiltonian (\ref{H}) is with exact $\mathcal{PT}$ symmetry if and only if $\abs{\eta}<\eta_c$, where $\eta_c:=\min\{2J,2J_s\}$.
That is, $M=\{(h,\eta)~|~h\in\RR, \abs{\eta}<\eta_c\}$.

Consider first the anisotropic case with $J\Gamma_s=J_s\Gamma$. From Eqs. (\ref{Energy}) and (\ref{Lambda}), we deduce that $\varepsilon_{n,k}$ is real if and only if $c_2(k)^2-c_4(k)\geq 0$ and $c_2(k)\geq 0$. Direct calculations show that
\begin{eqnarray}\label{eq:2-1}
c_2(k)^2-c_4(k)=4\left(-\eta^2+4J^2\cos(k)^2+4J_s^2\sin(k)^2\right)
\left(h^2+4\Gamma_s^2\cos(k)^2+4\Gamma^2\sin(k)^2\right)-
16(J\Gamma-J_s\Gamma_s)^2\sin(2k)^2.
\end{eqnarray}
Clear, a necessary condition for $c_2(k)^2-c_4(k)\geq 0$ is $(-\eta^2+4J^2\cos(k)^2+4J_s^2\sin(k)^2)\geq 0$ for all $k$, that is, $\abs{\eta}\leq\eta_c$. However, $\eta$ cannot equal to $\eta_c$, since the eigenstates of $D_k$ collapse at this point. So, $\abs{\eta}<\eta_c$. This proves the necessity of $\abs{\eta}<\eta_c$. To prove the sufficiency, we deduce from Eq. (\ref{eq:2-1}) that
\begin{eqnarray}\label{eq:2-2}
c_2(k)^2-c_4(k)&\geq & 16\left(-\eta^2+4J^2\cos(k)^2+4J_s^2\sin(k)^2\right)
\left(\Gamma_s^2\cos(k)^2+\Gamma^2\sin(k)^2\right)-
16(J\Gamma-J_s\Gamma_s)^2\sin(2k)^2\nonumber\\
&=&16\left[4\left(J\Gamma_s \cos(k)^2+J_s\Gamma\sin(k)^2\right)-\eta^2\left(\Gamma_s^2
\cos(k)^2+\Gamma^2\sin(k)^2\right)\right]\nonumber\\
&=&16\left[4J\Gamma_s-\eta^2\left(\Gamma_s^2
\cos(k)^2+\Gamma^2\sin(k)^2\right)\right]\nonumber\\
&=&16\left[4J_s\Gamma-\eta^2\left(\Gamma_s^2
\cos(k)^2+\Gamma^2\sin(k)^2\right)\right].
\end{eqnarray}
Here, at the second and third equalities, we have used the condition $J\Gamma_s=J_s\Gamma$. Let $f(k):=4J\Gamma_s-\eta^2(\Gamma_s^2
\cos(k)^2+\Gamma^2\sin(k)^2)=4J_s\Gamma-\eta^2(\Gamma_s^2
\cos(k)^2+\Gamma^2\sin(k)^2)$. If $\Gamma>\Gamma_s$, $\min_k f(k)=4J_s\Gamma-\eta^2\Gamma^2$. Under the condition $\abs{\eta}<\eta_c=\min\{2J,2J_s\}$, $\min_k f(k)> 0$.
If $\Gamma\leq\Gamma_s$, $\min_k f(k)=4J\Gamma_s-\eta^2\Gamma_s^2$. Similarly, $\min_k f(k)> 0$, too. Hence, $c_2(k)^2-c_4(k)> 0$. Besides, it is easy to see that $c_2(k)> 0$ when $\abs{\eta}<\eta_c$. Hence, all eigenvalues $\varepsilon_{n,k}$ are real. On the other hand, for any fixed $k$, the four eigenvalues $\varepsilon_{n,k}$, $n=1,2,3,4$, are distinct from each other, indicating that the corresponding eigenstates are complete. That is, the model is with exact $\mathcal{PT}$ symmetry. This proves the sufficiency.

Consider now the pseudo-isotropic case, i.e., $J\Gamma=J_s\Gamma_s$. In this case,
\begin{eqnarray}
c_2(k)^2-c_4(k)=4\left(-\eta^2+4J^2\cos(k)^2+4J_s^2\sin(k)^2\right)
\left(h^2+4\Gamma_s^2\cos(k)^2+4\Gamma^2\sin(k)^2\right).
\end{eqnarray}
Likewise, a necessary condition for $c_2(k)^2-c_4(k)\geq 0$ for all $k$ is $\abs{\eta}\leq\eta_c$. Again, $\abs{\eta}\neq\eta_c$. So, $\abs{\eta}<\eta_c$, thus proving the necessity. The sufficiency is obvious. Indeed, it is easy to see that under the condition $\abs{\eta}<\eta_c$, $c_2(k)^2-c_4(k)>0$ and $c_2(k)>0$. So, the eigenvalues $\varepsilon_{n,k}$ are real. Besides, duo to the same reason, the corresponding eigenstates are complete. That is, the model in this case is with exact $\mathcal{PT}$ symmetry, too.

\textit{Third, we find the critical points of the model.}
From Eqs. (\ref{Energy}) and (\ref{H-diag}), we deduce that
the GS of the model corresponds to the configuration that all the levels of negative energy, i.e., $\varepsilon_{n,k}=-\Lambda_\pm(k)$,
are occupied, whereas the levels of positive energy, i.e., $\varepsilon_{n,k}=\Lambda_\pm(k)$, are unoccupied.
As explained in the main text, critical points are of two types, i.e., the $\mathcal{PT}$PT points and QPT points. In the above paragraph, we have already found the $\mathcal{PT}$PT points, that is, for both cases, the $\mathcal{PT}$PT points are $(h,\eta)$ with $\abs{\eta}=\eta_c$. At these points, the eigenstates of $D_k$ collapse, resulting in drastic changes in the GS.

To find the QPT points, we need to discuss the two cases, respectively. Consider first the anisotropic case with $J\Gamma_s=J_s\Gamma$. After a moment of thought, one can easily figure out that the level crossing between the GS and excited states, i.e., the presence of the QPT points, occurs if and only if $c_4(k)=0$ for some $k$. Note that there are two non-negative terms, i.e., $[h^2+\eta^2-4(J^2-\Gamma_s^2)\cos(k)^2-
4(J_s^2-\Gamma^2)\sin(k)^2]^2$ and $16(J\Gamma-J_s\Gamma_s)^2\sin(2k)^2$, in $c_4(k)$. The two terms have to vanish simultaneously. Since $J\Gamma\neq J_s\Gamma_s$, the vanishing of the second term requires that $k=0$ or $k=\frac{\pi}{2}$. When $k=0$, the first term vanishes if and only if
\begin{eqnarray}\label{rc1}
r_{c,1}=2\sqrt{J^2-\Gamma_s^2},
\end{eqnarray}
and when $k=\frac{\pi}{2}$, it vanishes if and only if
\begin{eqnarray}\label{rc2}
r_{c,2}=2\sqrt{J_s^2-\Gamma^2}.
\end{eqnarray}
Here, $r:=\sqrt{h^2+\eta^2}$ denotes the modulus of the the field strength, and the relations $J>\Gamma_s$ and $J_s>\Gamma$ are assumed. Therefore, the QPT points are the $(h,\eta)$ with Eq. (\ref{rc1}) and those $(h,\eta)$ with Eq. (\ref{rc2}).
Consider now the pseudo-isotropic case. Again, the QPT points are the points such that $c_4(k)=0$ for some $k$. Note that $c_4(k)=[h^2+\eta^2-4(J^2-\Gamma_s^2)\cos(k)^2-
4(J_s^2-\Gamma^2)\sin(k)^2
]^2$ in this case. $c_4(k)=0$ for some $k$ if and only if $r$ lies between $r_{c,1}$ and $r_{c,2}$, i.e., $\min\{r_{c,1},r_{c,2}\}\leq r\leq\max\{r_{c,1},r_{c,2}\}$.
So, the QPT points are the $(h,\eta)$ such that $\min\{r_{c,1},r_{c,2}\}\leq r\leq\max\{r_{c,1},r_{c,2}\}$. Now, we have identified the critical points analytically. The results are shown in Fig. \ref{fig}.

\textit{Finally, we exploit the metric tensor to identify the critical points and further confirm its usefulness.} To this end, we need to find the form of the GS. Let us introduce a reference state $\ket{ref}$, defined as
\begin{eqnarray}\label{ref}
\tilde{b}_{n,k}\ket{ref}=0, ~~~~\forall n, k.
\end{eqnarray}
From Eq. (\ref{b-b}) and noting that $\bm{a}_k=(a_k,a_{-k}^\dagger,a_{\pi-k}^\dagger,a_{k-\pi})$, we deduce that
\begin{eqnarray}
\ket{ref}=\ket{0_k,1_{-k},1_{\pi-k},0_{k-\pi}},
\end{eqnarray}
where $\ket{0_k}$ denotes the vacuum of the $k$-th mode, and $\ket{1_{-k}}$, $\ket{1_{\pi-k}}$, and $\ket{0_{k-\pi}}$ are defined in a similar manner.
Using Eq. (\ref{ref}) and noting that $[b_{n,k}^\dagger\tilde{b}_{n,k},b_{n^\prime,k^\prime}^\dagger]=
\delta_{nn^\prime}\delta_{kk^\prime}b_{n^\prime,k^\prime}^\dagger$, we have
\begin{eqnarray}
Hb_{n_1,k_1}^\dagger b_{n_2,k_2}^\dagger\cdots\ket{ref}
=(\varepsilon_{n_1,k_1}+\varepsilon_{n_2,k_2}+\cdots)
b_{n_1,k_1}^\dagger b_{n_2,k_2}^\dagger\cdots\ket{ref}.
\end{eqnarray}
So, the form of the GS of $H$ reads
\begin{eqnarray}\label{GS1}
&&\ket{\Psi_{0,\lambda}}=\prod_{\varepsilon_{n,k}<0}b_{n,k}^\dagger
\ket{ref}=\nonumber\\
&&\prod_{\varepsilon_{n,k}<0}\left(\psi_{n,k}^1
\ket{1_k,1_{-k},1_{\pi-k},0_{k-\pi}}
+\psi_{n,k}^2
\ket{0_k,0_{-k},1_{\pi-k},0_{k-\pi}}
+\psi_{n,k}^3
\ket{0_k,1_{-k},0_{\pi-k},0_{k-\pi}}
+\psi_{n,k}^4
\ket{0_k,1_{-k},1_{\pi-k},1_{k-\pi}}\right).\nonumber\\
\end{eqnarray}
Here, $\psi_{n,k}^i$, $i=1,2,3,4$, denote the components of $\ket{\psi_{n,k}}$, i.e., $\ket{\psi_{n,k}}=(\psi_{n,k}^1,
\psi_{n,k}^2,\psi_{n,k}^3,\psi_{n,k}^4)^T$. Similar analysis shows that the form of the GS of $H^\dagger$ reads
\begin{eqnarray}\label{GS2}
&&\ket{\Phi_{0,\lambda}}=\prod_{\varepsilon_{n,k}<0}\tilde{b}_{n,k}^\dagger
\ket{ref}=\nonumber\\
&&\prod_{\varepsilon_{n,k}<0}\left(\phi_{n,k}^1
\ket{1_k,1_{-k},1_{\pi-k},0_{k-\pi}}
+\phi_{n,k}^2
\ket{0_k,0_{-k},1_{\pi-k},0_{k-\pi}}
+\phi_{n,k}^3
\ket{0_k,1_{-k},0_{\pi-k},0_{k-\pi}}
+\phi_{n,k}^4
\ket{0_k,1_{-k},1_{\pi-k},1_{k-\pi}}\right),\nonumber\\
\end{eqnarray}
where $\phi_{n,k}^i$, $i=1,2,3,4$, denote the components of $\ket{\phi_{n,k}}$, i.e., $\ket{\phi_{n,k}}=(\phi_{n,k}^1,
\phi_{n,k}^2,\phi_{n,k}^3,\phi_{n,k}^4)^T$. Substituting Eqs.~(\ref{GS1}) and (\ref{GS2}) into Eq.~(9) in the main text, we obtain
\begin{eqnarray}
g_{0,\mu\nu}=\frac{1}{2}\Re\sum_{\varepsilon_{n,k}<0}
\inp{\partial_\mu\phi_{n,k}}{\partial_\nu\psi_{n,k}}
-\inp{\partial_\mu\phi_{n,k}}{\psi_{n,k}}
\inp{\phi_{n,k}}{\partial_\nu\psi_{n,k}}
+\inp{\partial_\mu\psi_{n,k}}{\partial_\nu\phi_{n,k}}
-\inp{\partial_\mu\psi_{n,k}}{\phi_{n,k}}
\inp{\psi_{n,k}}{\partial_\nu\phi_{n,k}}.\nonumber\\
\end{eqnarray}
In the thermodynamic limit, i.e., $L\rightarrow\infty$, one can replace the discrete variable $k$ with a continuous variable and substitute the sum with an integral, resulting in
\begin{eqnarray}
&&g_{0,\mu\nu}=\nonumber\\
&&\frac{L}{4\pi}\Re\sum_{\textrm{occupied}~n}\int_{0}^{\pi/2}
\ud k
\inp{\partial_\mu\phi_{n,k}}{\partial_\nu\psi_{n,k}}
-\inp{\partial_\mu\phi_{n,k}}{\psi_{n,k}}
\inp{\phi_{n,k}}{\partial_\nu\psi_{n,k}}
+\inp{\partial_\mu\psi_{n,k}}{\partial_\nu\phi_{n,k}}
-\inp{\partial_\mu\psi_{n,k}}{\phi_{n,k}}
\inp{\psi_{n,k}}{\partial_\nu\phi_{n,k}}.\nonumber\\
\end{eqnarray}
In the thermodynamic limit, the intensity of the metric tensor, i.e., $\bar{g}_{\mu\nu}:=\lim_{L\rightarrow\infty}g_{0,\mu\nu}/L$, is of interest, which is analogies to the fact that the intensity of the free energy rather than the free energy itself is of interest for QPTs of Hermitian systems \cite{1975Perk319}. It is easy to see that the intensity of the metric tensor reads
\begin{eqnarray}\label{Intensity}
&&\bar{g}_{\mu\nu}=\nonumber\\
&&\frac{1}{4\pi}\Re\sum_{\textrm{occupied}~n}\int_{0}^{\pi/2}
\ud k
\inp{\partial_\mu\phi_{n,k}}{\partial_\nu\psi_{n,k}}
-\inp{\partial_\mu\phi_{n,k}}{\psi_{n,k}}
\inp{\phi_{n,k}}{\partial_\nu\psi_{n,k}}
+\inp{\partial_\mu\psi_{n,k}}{\partial_\nu\phi_{n,k}}
-\inp{\partial_\mu\psi_{n,k}}{\phi_{n,k}}
\inp{\psi_{n,k}}{\partial_\nu\phi_{n,k}}.\nonumber\\
\end{eqnarray}
The singularities of $\bar{g}_{\mu\nu}$ represent the critical points, as explained in the main text. Now, what we need to do is to calculate $\bar{g}_{\mu\nu}$ for each $(h,\eta)\in M$.
Due to the complexity of the expressions of $\ket{\psi_{n,k}}$
and $\ket{\phi_{n,k}}$, it is demanding to perform the integral in Eq. (\ref{Intensity}) analytically, but it is very easy to calculate it numerically. The numerical results are shown in Fig. 1 of the main text. Comparing Fig. 1 of the main text and Fig. \ref{fig} here (see also the explanation in the main text), one can find that the metric tensor exhibit singular behavior at $(h,\eta)$ if and only if $(h,\eta)$ is one of the critical points identified analytically in the previous paragraphs. This confirms the usefulness of the metric tensor in identifying critical points.

\end{document}